\documentclass[manuscript,screen]{acmart}
\usepackage{multicol, multirow}
\usepackage{xspace}
\usepackage{algorithm2e}
\usepackage{booktabs}
\usepackage{algorithmic}
\usepackage{subfigure}
\usepackage{tablefootnote}

\newcommand{\eat}[1]{}
\newcommand{\ie}{\emph{i.e.,}\xspace}
\newcommand{\eg}{\emph{e.g.,}\xspace}
\newcommand{\etal}{\emph{et al.}\xspace}
\newcommand{\paratitle}[1]{\smallskip\noindent \textbf{#1}}
\AtBeginDocument{%
  }

\setcopyright{acmlicensed}
\copyrightyear{2018}
\acmYear{2018}
\acmDOI{XXXXXXX.XXXXXXX}
\acmConference[Conference acronym 'XX]{Make sure to enter the correct conference title from your rights confirmation email}{June 03--05,2018}{Woodstock, NY}

\acmISBN{978-1-4503-XXXX-X/2018/06}

\begin{document}

\title{Multi-Interest Recommendation: A Survey}

\author{Zihao Li}
\email{zihao.li@whu.edu.cn}
\affiliation{%
  \institution{Key Laboratory of Aerospace Information Security and Trusted Computing, Ministry of Education, School of Cyber Science and Engineering, Wuhan University}
  \city{Wuhan}
  \country{China}
}

\author{Qiang Chen}
\affiliation{%
  \institution{Tencent WeChat}
  \city{Guangzhou}
  \country{China}}
\email{ethanqchen@tencent.com}



\author{Lixin Zou}
\email{zoulixin@whu.edu.cn}
\affiliation{%
  \institution{Key Laboratory of Aerospace Information Security and Trusted Computing, Ministry of Education, School of Cyber Science and Engineering, Wuhan University}
  \city{Wuhan}
  \country{China}
}

\author{Aixin Sun}
\affiliation{%
  \institution{Nanyang Technological University}
  \city{Singapore}
  \country{Singapore}}
\email{axsun@ntu.edu.sg}

\author{Chenliang Li}
\authornote{Chenliang Li is the corresponding author.}
\email{cllee@whu.edu.cn}
\affiliation{
\institution{Key Laboratory of Aerospace Information Security and Trusted Computing, Ministry of Education, School of Cyber Science and Engineering, Wuhan University}
\city{Wuhan}
\country{China}
}
\renewcommand{\shortauthors}{Li et al.}

\begin{abstract} 
  Existing recommendation methods often struggle to model users' multifaceted preferences due to the diversity and volatility of user behavior, as well as the inherent uncertainty and ambiguity of item attributes in practical scenarios.
  Multi-interest recommendation addresses this challenge by extracting multiple interest representations from users’ historical interactions, enabling fine-grained preference modeling and more accurate recommendations.
  It has drawn broad interest in recommendation research. However, current recommendation surveys have either specialized in frontier recommendation methods or delved into specific tasks and downstream applications.
 In this work, we systematically review the progress, solutions, challenges, and future directions of multi-interest recommendation by answering the following three questions: (1) \textbf{\textit{Why}} is multi-interest modeling significantly important for recommendation? (2) \textbf{\textit{What}} aspects are focused on by multi-interest modeling in recommendation? and (3) \textbf{\textit{How}} can multi-interest modeling be applied, along with the technical details of the representative modules?
 We hope that this survey establishes a fundamental framework and delivers a preliminary overview for researchers interested in this field and committed to further exploration. 
The implementation of multi-interest recommendation summarized in this survey is maintained at \url{https://github.com/WHUIR/Multi-Interest-Recommendation-A-Survey}.
 
\end{abstract}

\ccsdesc[100]{Information systems~Recommender systems}

\keywords{Multi Interest, Multi Preference, Interest Modeling, Recommendation}


\maketitle

\section{Introduction}
\label{sec:intro}

\begin{figure}[h!]
\centering
\includegraphics[width=\columnwidth]{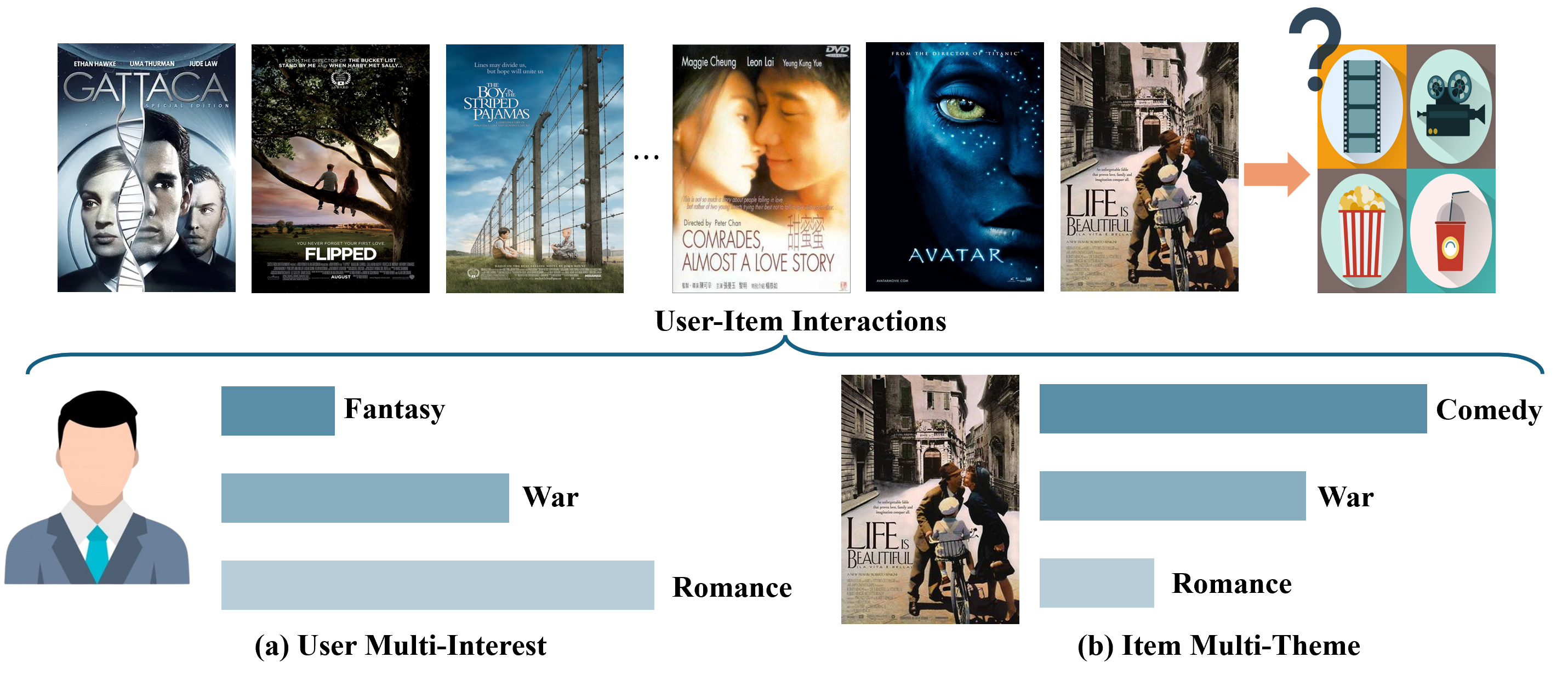}
\caption{A toy example of movie recommendation, where a user interacts with movies spanning multiple themes or genres, revealing her multiple interests. Also, each movie encompasses several genres and themes.  
}
\label{fig:toy_multi}
\end{figure}

Recommendation system~\cite{resnick1997recommender} aims to filter and suggest information or products based on user historical interaction records, assisting users in making decisions more efficiently.
Over the past decades, it has achieved remarkable success across various domains, including e-commerce~\cite{schafer1999recommender}, social media~\cite{guy2010social}, news~\cite{karimi2018news}, entertainment~\cite{harper2015movielens}, and others.
However, user preferences and behavior patterns are becoming more complex and uncertain, given the surge in products and services. 
It presents a new challenge to the frontier of recommendation systems.
Exemplifier by a movie recommendation scenario (as shown in Figure~\ref{fig:toy_multi}), a user is interested in different genres (\eg romance, fiction, comedy, etc), simultaneously a movie also belong to more than one genre (\eg \textit{La vita è bella} (Life is Beautiful) is categorized as romance, comedy, and war). 
General recommendation methods often struggle to capture the diverse and multifaceted interests of users effectively when tackling such intricate user interactions and item themes.
To address this problem, multi-interest recommendation (also known as multi-faceted preference recommendation), endeavoring to model users' multifaceted and diverse interests at a finer granularity for recommendation enhancement~\cite {li2019multi}, has emerged as a prevalent approach.

\begin{figure}[]
\centering
\includegraphics[width=\columnwidth]{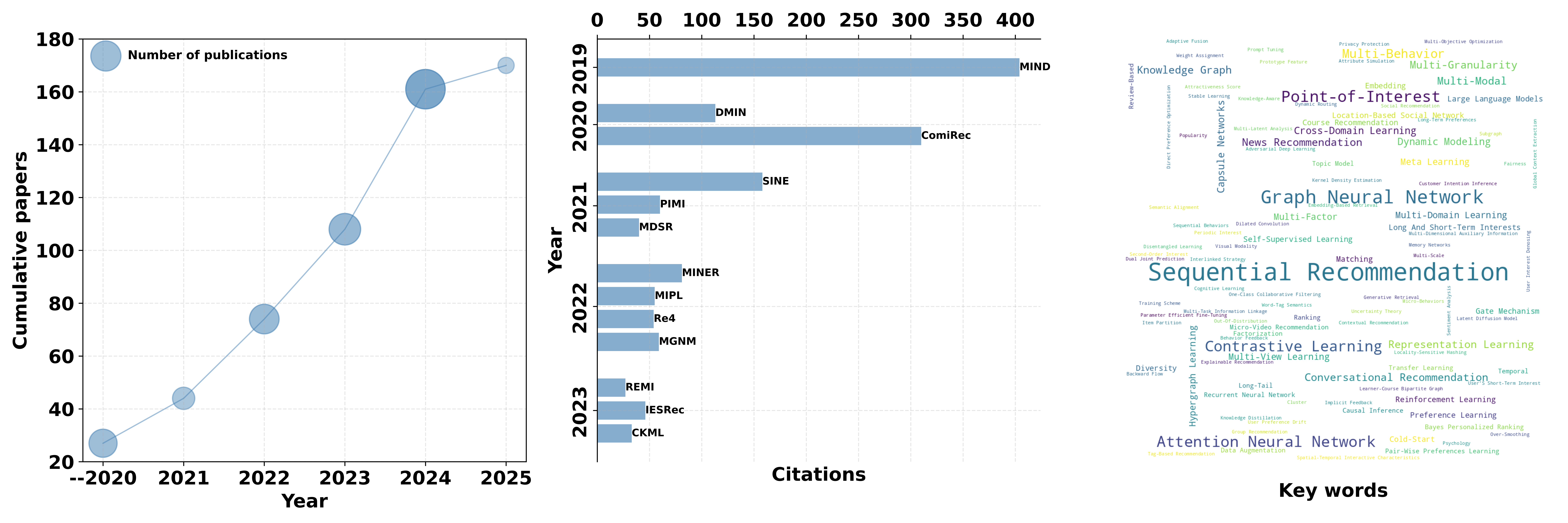}
\caption{The left line chart illustrates the cumulative number of multi-interest recommendation publications, retrieved from the DBLP platform. The related works increased year by year and reached a total of $172$ by March 4, 2025. The size of the circles is proportional to the number of publications for the given year. The middle bar presents the representative works and their citations. The word cloud on the right displays the frequency of keywords extracted from the $172$ papers.
}
\label{fig:count}
\end{figure}

Reviewing the research of multi-interest recommendation, Li \etal~\cite{li2005hybrid} first proposed the concept of multiple interest (\ie \textit{users have many completely different interests}) in e-commerce recommendation in $2005$.
In 2007, Wang~\cite{wang2007multi} modeled users' multi-interests for community-based recommendation. 
However, this research direction did not receive significant attention in its early stages. 
In $2019$, MIND~\cite{li2019multi} generated multiple vectors to represent the user's interests by using a dynamic routing mechanism. 
This work comprehensively demonstrated the effectiveness of multi-interest recommendation in processing major online traffic on the homepage of the Mobile Tmall App~\footnote{The largest Business-To-Customer (B2C) e-commerce platform (\url{https://www.tmall.com/}) in China boasts hundreds of millions of active users and offers millions of products spanning various categories.}.
Since then, a growing number of researchers have delved into this topic, resulting in the successive emergence of influential works, including ComiRec~\cite{cen2020controllable}, SINE~\cite{tan2021sparse}, MINER~\cite{li2022miner}, and so on. 
Multi-interest recommendation has become a hot topic~\cite{isinkaye2015recommendation} and has attracted broad attention from academia and industry.
Figure~\ref{fig:count} (middle subfigure) presents the development timeline and milestones in this research area. All of these works are published in top venues, including KDD, Web Conference, SIGIR, WSDM, CIKM, RecSys, ACL, IJCAI, and TOIS.   
The left line chart in Figure~\ref{fig:count} illustrates that the number of papers with regard to multi-interest recommendation has been growing (\ie from $27$ before 2021 to $53$ in 2024), reaching a total of $172$ papers by March 4, 2025\footnote{We count the published works of the multi-interest and multi-preference recommendation retrieved from DBLP (\url{https://dblp.org/search}) platform by searching for titles containing the phrases "multi-interest recommendation" or "multi-preference recommendation".}.

\begin{figure}[]
\centering
\includegraphics[width=\columnwidth]{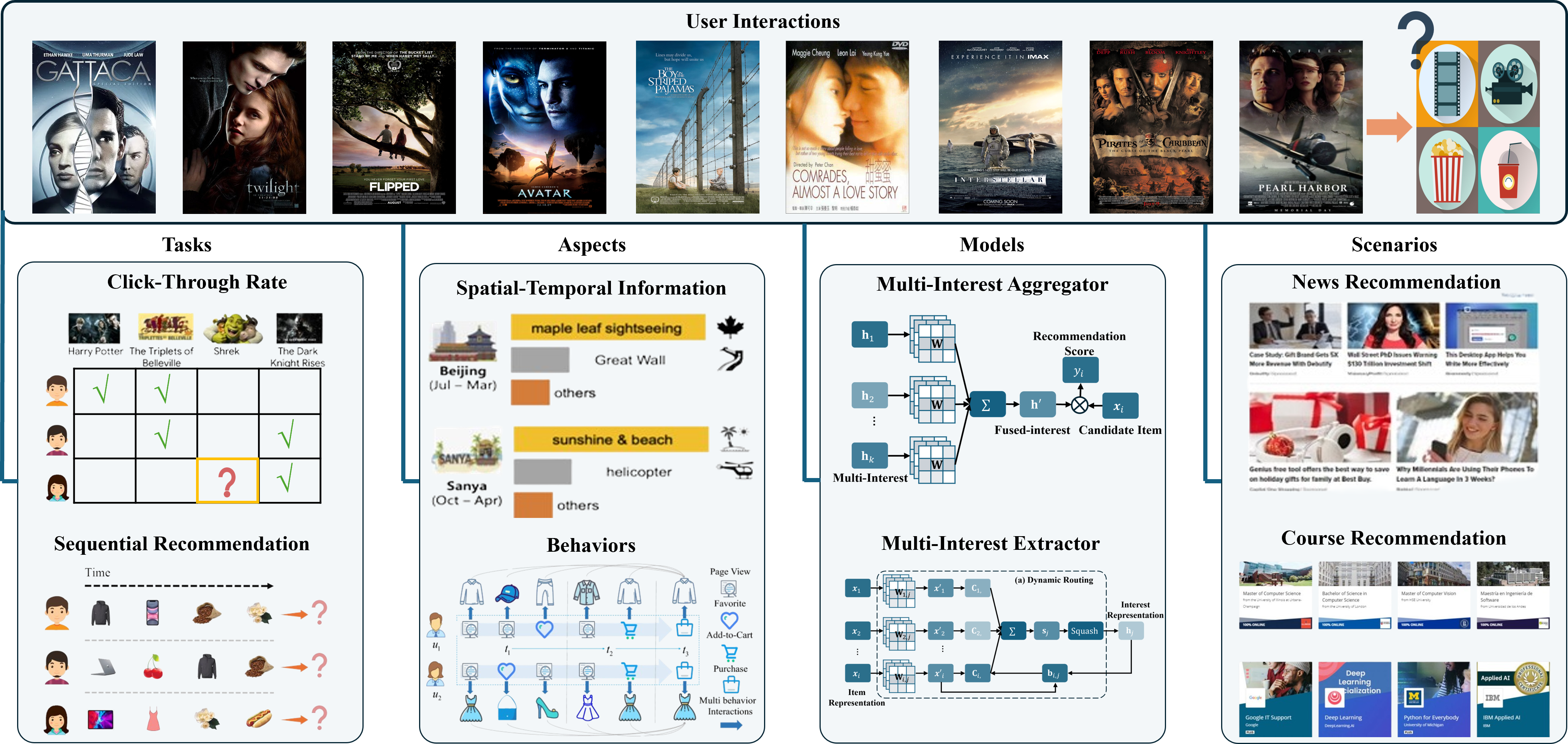}
\caption{The toy example of a user's multi-interest interaction history, as depicted in the top diagram. The diagrams below depict the key research dimensions of multi-interest recommendations, including recommendation tasks, multi-interest modeling aspects, key components, and application scenarios. 
}
\label{fig:toy}
\end{figure}

We quantify keyword frequencies from published works to further contextualize current research trends in this research area, as illustrated in Figure~\ref{fig:count} (right). Analyzing the word cloud, the key research dimensions within this domain can be categorized into the five aspects: tasks, modeling aspects, methodologies, application scenarios, and future directions, as shown in Figure~\ref{fig:toy}.
Specifically, \textbf{(1) Tasks:} sequential recommendation~\cite{quadrana2018sequence}, specializing in users' interest evolution and dynamic shifts, dominates focus and efforts in multi-interest recommendation. Click-through rate prediction~\cite{su2009survey} follows as a secondary research task.
Additionally, conversational recommendation~\cite{zheng2024facetcrs} is increasingly explored in multi-interest modeling due to the capture of users' real-time intentions.
\textbf{(2) Modeling Aspects:} the focus aspects of multi-interest including user behaviors~\cite{li2024multi}, item attributes and topics~\cite{li2015learning}, domain information~\cite{jiang2022adaptive}, spatial-temporal information~\cite{tao2022sminet} and multi-modal information~\cite{zhao2023m5}. 
\textbf{(3) Methodologies:} a multi-interest extractor and a multi-interest aggregator are two main components in a typical multi-interest recommendation framework. Concretely, the multi-interest extractor endeavors to obtain the multiple interest representation by dynamic routing~\cite{sabour2017dynamic, li2019capsule}, attention mechanism, and so on. The multi-interest aggregator then fuses these multiple interest representations, or the outputs yielded by each, for the final recommendation.
Graph neural networks (GNNs)~\cite{scarselli2008graph} and transformer-based architectures~\cite{vaswani2017attention} are the predominant model architecture in multi-interest recommendation, owing to the remarkable ability in auxiliary information aggregation and sequential pattern modeling. 
Complementary techniques include knowledge graphs~\cite{wang2014knowledge}, meta-learning~\cite{finn2017model}, and contrastive learning~\cite{chen2020simple} are also explored for multi-interest recommendation.
\textbf{(4) Application Scenarios:} e-commence~\cite{wei2007survey}, micro-video~\cite{li2022improving}, movies~\cite{goyani2020review}, news~\cite{li2019multi}, and course recommendations~\cite{wang2024mooc} emerge as practical applications, garnering specific attention in multi-interest recommendation. 
\textbf{(5) Future Directions:} despite promising results in this research area, several research lines do not receive sufficient investigation and exploration, such as large language models-driven solutions and their associated directions, \eg preference alignment, parameter-efficient fine-tuning, and knowledge distillation~\cite{qiao2024llm, bai2024aligning}; diffusion methods~\cite{le2025diffusion}; reinforcement learning frameworks~\cite{zhang2022multiple}; long-tail and cold start challenges~\cite{zhang2022diverse, liu2023co}; fairness and security in recommendation systems~\cite{zhao2024can, meng2023intelligent}.
The subsequent sections of this paper provide a comprehensive analysis of these dimensions.

\paratitle{The differences between this survey and former ones.} The existing recommendation system surveys published in top venues are organized in Table~\ref{tab:surveys}.
In general, these reviews can be roughly divided into five groups based on their main topics: comprehensive surveys (\ie diving into recommendation systems as a whole, aiming to provide a holistic overview and a systematic mapping of this field), recommendation tasks, key methodologies, application scenarios, and others.
It is noted that these surveys encompass a large spectrum of topics, including both burgeoning cutting-edge (\eg large language models and diffusion methods) and specialized directions (\eg cloud and financial service, privacy-preserving and trustworthy in recommendation).
Although in~\cite{jannach2021survey} the authors provide a comprehensive survey regarding intent-aware recommendation systems, it is significantly different than multi-interest modeling. Specifically, intent modeling aims to \textit{capture the users’ current underlying motivations for recommendations}. In contrast, multi-interest modeling attempts to \textit{capture and extract the user's multiple and diverse interests for recommendation enhancement}.   
Therefore, none of the existing surveys are dedicated to multi-interest modeling, despite its critical role in advancing fine-grained user representation and accurate recommendations.
To bridge this gap, we provide a comprehensive overview of the multi-interest recommendation in this survey.

\begin{table}[]
\caption{Summary of surveys in recommendation system. It can be categorized into five dimensions: comprehensive survey, task-oriented, application-oriented, methodology-oriented, and other hot topics.}
\tabcolsep=0.07cm
\small
\begin{tabular}{lll||lll}
\hline\hline
\multicolumn{2}{l}{\textbf{Comprehensive Survey}}                                & \cite{da2020recommendation, lu2015recommender, bobadilla2013recommender, zhang2019deep, burke2002hybrid, adomavicius2005toward, roy2022systematic} &  \multirow{10}{*}{\rotatebox[origin=c]{90}{\textbf{Application}}} & Music                               & \cite{schedl2019music, kaminskas2012contextual, aaman2010survey, deldjoo2024content}                                            \\ \cline{1-3}
\multirow{7}{*}{\rotatebox[origin=c]{90}{\textbf{Tasks}}}    & Sequential Recommendation            & \cite{boka2024survey, quadrana2018sequence, chen2023survey, fang2020deep}                                                                          &                                 & News                                & \cite{karimi2018news, raza2022news, meng2023survey}                                                                             \\
                          & Session-Based Recommendation         & \cite{wang2021survey, li2024graph}                                                                                                                 &                                & Course \& Education                 & \cite{da2023systematic, liu2022review}                                                                                          \\
                          & Click-Through Rate Prediction        & \cite{yang2022click, zhang2021deep}                                                                                                                &                               & E-Recruitment                       & \cite{mashayekhi2024challenge, freire2021recruitment}                                                                           \\
                          & Point-of-Interest Recommendation     & \cite{yu2015survey, zhao2016survey, islam2022survey}                                                                                               &                              & Literature \& Academic                & \cite{ali2020deep,champiri2015systematic, beel2016paper}                                                                        \\
                          & Cross-Domain Recommendation          & \cite{zang2022survey, khan2017cross, fernandez2012cross}                                                                                           &                               & Tourism                             & \cite{borras2014intelligent,banerjee2023review}                                                                                 \\
                          & Group Recommendation                 & \cite{alvarado2022systematic, kompan2014group}                                                                                                     &                                & Cloud Services \& Financial Services & \cite{zibriczky122016recommender, aznoli2017cloud}                                                                              \\
                          & Bundle Recommendation                & \cite{sun2024survey, sun2022revisiting}                                                                                                            &                                & Health Care \& Health Food          & \cite{ali2023deep, trang2018overview, elsweiler2012food}                                                                        \\ \cline{1-3}
\multirow{13}{*}{\rotatebox[origin=c]{90}{\textbf{Methods}}} & Collaborative Filtering              & \cite{Desrosiers2011neig, koren2021advances, shi2014collaborative, elahi2016survey}                                                                &                             & Social Network \& Social Medial     & \cite{li2023survey, yu2015survey, sharma2024survey}                                                                             \\
                          & Matrix Completion                    & \cite{ramlatchan2018survey, bokde2015matrix}                                                                                                       &                                 & E-commerce \& Fashion               & \cite{wei2007survey,ding2023computational, deldjoo2023review, gong2021aesthetics}                                               \\ \cline{4-6} 
                          & Graph Neural Networks                & \cite{li2023survey, li2024graph, wu2022graph, gao2023survey, anand2025survey}                                                                      &  \multirow{11}{*}{\rotatebox[origin=c]{90}{\textbf{Hot Topic}}}   & Explainable Recommendation           & \cite{zhang2020expr,chatti2024visualization, chen2022measuring, zhang2020explainable}                                           \\
                          & LLMs-Based                           & \cite{wu2024llms, liu2023pre, bao2023large, zhao2024recommender, lin2025can}                                                                       &                             & Evaluation                          & \cite{zhang2019eval, gunawardana2009survey, zangerle2022evaluating}                                                             \\
                          & Diffusion Method                     & \cite{lin2024survey, wei2025diffusion}                                                                                                             &                             & Data Sparsity \& Cold Start Problem & \cite{chen2024data, singh2020scalability, jiang2022short}                                                                       \\
                          & Reinforcement Learning              & \cite{afsar2022reinforcement, lin2023survey, chen2023deep, chen2024opportunities, silva2022multi}                                                  &                             & Fairness                            & \cite{li2023fairness, jin2023survey, wang2023survey, zhao2025fairness, wu2023fairness,zehlike2022fairness, vassoy2024consumer} \\
                          & Knowledge Graph-Based                & \cite{tarus2018knowledge, guo2020survey, shao2021survey, zhang2024review}                                                                          &                             & Diversity \& Long-Tail Problem      & \cite{wu2024result, kaminskas2016diversity, kunaver2017diversity}                                                               \\
                          & Multi-Modal \& Multi-Behavior         & \cite{liu2024multimodal, liu2024multimodala, chen2023survey}                                                                                       &                             & Debiasing \& Denoising \& Sampling     & \cite{gupta2024unbiased, chen2023bias, jain2023sampling, ma2024negative}                                                        \\
                          & Causal Inference                     & \cite{luo2024survey, xu2023causal, gao2024causal}                                                                                                  &                             & Federated Recommendation            & \cite{sun2024survey, javeed2023federated, wang2024horizontal}                                                                   \\
                          & Learning-to-Rank                     & \cite{karatzoglou2013learning, gupta2024unbiased, huang2016survey}                                                                                 &                             & Privacy-Preserving \& Trustworthy   & \cite{wang2018toward, xu2016privacy,ge2024survey, wang2024trustworthy, dong2022survey, liu2020long}                             \\
                          & Self-Supervised Learning             & \cite{ren2024comprehensive, yu2023self, jing2023contrastive}                                                                                       &                             & Context-Aware Recommendation        & \cite{kulkarni2020context, verbert2012context, raza2019progress, mateos2025systematic, champiri2015systematic}                  \\
                          & Active Learning \& Transfer Learning & \cite{elahi2016survey, rubens2015active, pan2016survey}                                                                                            &                             & Conversational Recommendation       & \cite{jannach2021survey, fu2020tutorial, lei2020conversational}                                                                 \\
                          & Meta Learning                        & \cite{wang2022deep, jiang2022short, cunha2018metalearning}                                                                                         &                          &   Intent-Aware Recommendation                                     &   \cite{jannach2024survey}                                                                                                                                               \\ \hline\hline
\end{tabular}
\label{tab:surveys}
\end{table}

\paratitle{Contributions of this survey}.To sum up, compared to existing surveys, the main contributions and unique features of this work are summarized as follows:
\textbf{(1) Delving into Multi-interest Modeling}. This is the first work to comprehensively summarize the existing research regarding user multi-interest modeling in recommendation systems, highlighting the advantages and significance of this research area to the broader community.
\textbf{(2) Innovative Categorization for Multi-interest Modeling Methods}. We introduce a systematic framework for multi-interest modeling, in which we classify the research works from three primary aspects: recommendation tasks, multi-interest modeling aspects, and motivations. In addition, we formulate a framework that allows us to structure and analyze the relevant methods.
\textbf{(3) Identification of Limitations and Future Research Directions}. This paper identifies the challenges and limitations in existing works that remain unresolved. Furthermore, we outline prospective research directions that are valuable for in-depth investigation.
The remaining content of this survey is arranged as follows: Section~\ref{sec:preliminary} provides a detailed introduction to the key concepts and significance of multi-interest modeling (\ie \textit{why multi-interest modeling}) first.
Then, we summarize the aspects of multi-interest modeling (\ie \textit{what are the aspects of multi-interest modeling}) from the user and item perspectives. Hence, the existing works can be systematically classified accordingly.
Section~\ref{sec:method} first formalizes the definition of multi-interest modeling in the recommendation tasks, including sequential recommendation and click-through prediction.
After that, we present the overall framework of multi-interest recommendation. Specifically, we delve into two typical components: multi-interest extractor and multi-interest aggregator, providing a detailed introduction to the representative techniques employed in these components (\ie \textit{how to multi-interest modeling}). 
In Section~\ref{sec:application}, we examine the specific practical applications and scenarios of multi-interest recommendation, such as movie and micro-video recommendation, news recommendation, life services, and travel recommendation.
Section~\ref{sec:challenges} discusses the challenges and future directions, including advancements in large-language models, adaptive and efficient multi-interest modeling, the long-tail and cold start recommendation problems in this research area. Finally, Section~\ref{sec:conclu} summarizes the key contributions of this survey.

\section{Motivations, and Aspects of Multi-Interest Recommendation}
\label{sec:preliminary}

In this section, we first summarize the different aspects of multi-interest modeling, and the corresponding concepts are then briefly introduced. Besides, we discuss the significance and necessity of multi-interest modeling in the recommendation system.
Accordingly, a systematic classification of representative works in this research domain across three dimensions: tasks, aspects of multi-interest modeling, and underlying motivations, is provided.

\eat{
\begin{table}[h!]
\caption{Key notations in this paper.}
\begin{tabular}{ll}
\hline\hline
\multicolumn{1}{c|}{Notations} & Descriptions     \\ \hline\hline
\multicolumn{1}{c|}{$\mathcal{U}$}         & The set of users \\
\multicolumn{1}{c|}{$\mathcal{I}$}         & The set of items                 \\
\multicolumn{1}{c|}{$\mathbf{R}$}          & User-item interaction matrix                 \\
\multicolumn{1}{c|}{$\hat{y}$}          & Predicted recommendation score                 \\
\multicolumn{1}{c|}{$\mathbf{x}$}          & Item representation                 \\
\multicolumn{1}{c|}{$\mathbf{h}$}          & Interest representation                 \\
\multicolumn{1}{c|}{$\mathbf{W}$}        & Learnable parameters            \\
\multicolumn{1}{c|}{$\mathcal{L}$}        & Loss function            \\
\hline\hline
\end{tabular}
\label{tab:notations}
\end{table}
}

\subsection{Motivation of Multi-Interest Recommendation}
\label{sec:why}

As introduced above, the primary motivation of multi-interest recommendation aims to model the users' multiple preferences and items' multifaceted aspects in practical scenarios.  
Specifically, the motivations of existing works in multi-interest recommendation can be summarized below:

\paratitle{Fine-grained Modeling of Users' Diverse Preferences.} User preferences are inherently diverse and uncertain, which will dynamically evolve across the temporal, spatial, social network, and domains~\cite{tao2022sminet, jiang2022adaptive, jiang2020aspect}. An advanced recommendation system is able to exploit the user's multi-level and multifaceted preferences, modeling users' multi-interests in a fine-grained manner, for an accurate recommendation. 

\paratitle{Fine-grained Modeling of Items' Multiple Aspects.} Akin to the user's multi-preferences, items also encompass multi-aspect, including attributes, themes, categories, multi-modality information, and others~\cite{pei2024rimirec, zheng2024facetcrs, li2019multi}. Users may be attracted to various facets of items. Hence, it is feasible and necessary to model users' multi-interests in alignment with the heterogeneous aspects of items, rather than focusing on an isolated dimension of items for recommendation.

\paratitle{Enhanced the Diversity of Recommendations.} Compared to modeling a user's interest from a single aspect, multi-interest recommendation allows a comprehensive exploration of user preferences across multiple dimensions. Therefore, by accounting for these multiple-interest points, the recommender system is capable of generating a diverse and wider array of recommendations, enriching the recommendation outcomes, and aligning with the dynamics of user preferences. 

\paratitle{Enhanced the Explainability of Recommendation.} Owing to the fine-grained modeling of the users' multi-interest and items' multi-aspects, the recommendation system can more accurately discern what the user truly prefers and which aspects of items are most appealing to the user. Therefore, it facilitates the development of explainable recommendations.

\subsection{Aspects of Multi-Interest Modeling}
\label{sec:diemnsion}  

\begin{figure}[]
\centering
\includegraphics[width=\columnwidth]{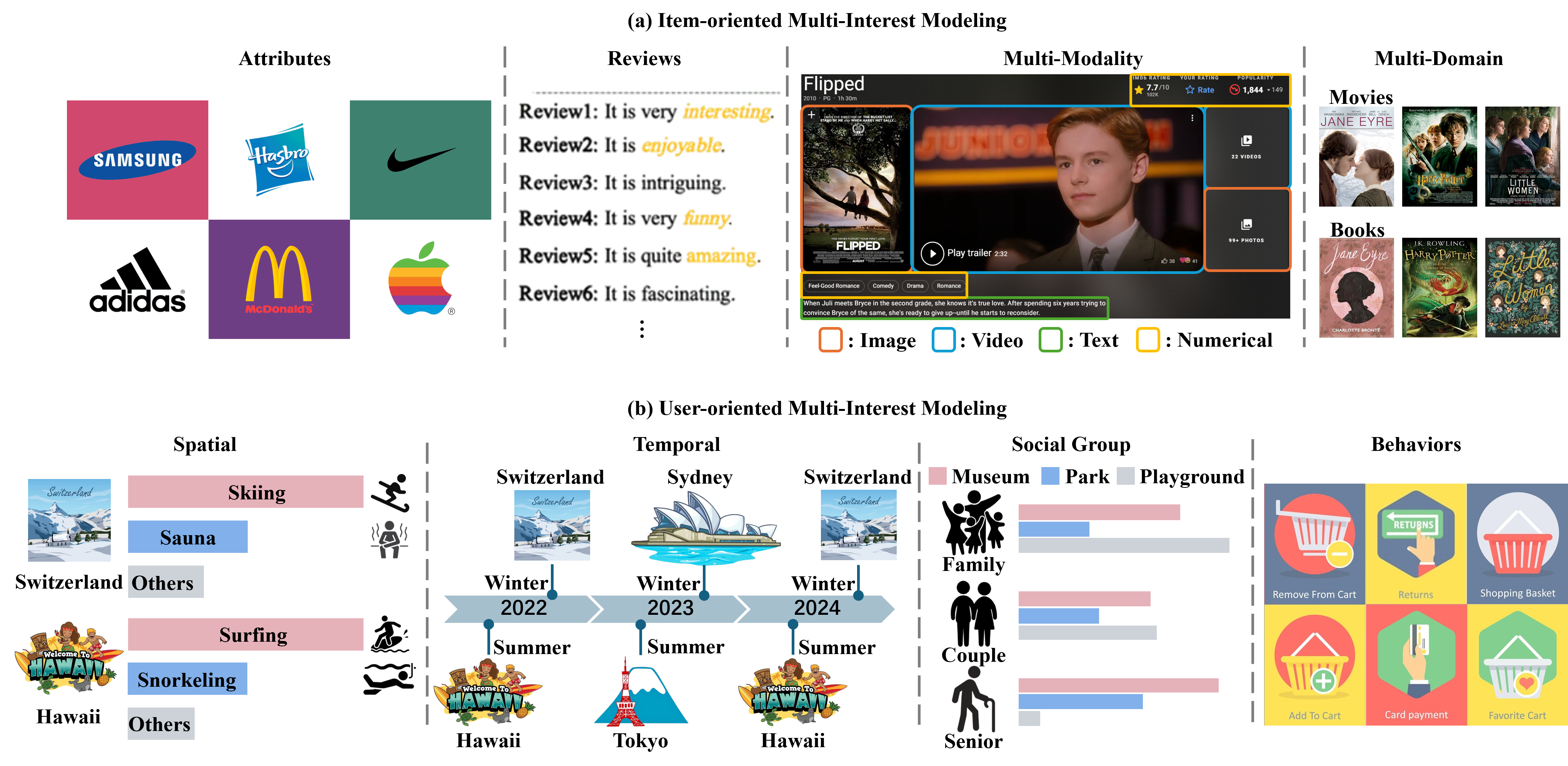}
\caption{The diagram of explicit multi-interest modeling aspects. It can be divided into item-oriented multi-interest modeling (top) and user-oriented multi-interest modeling (bottom).}
\label{fig:interest_aspect}
\end{figure}

To sum up, the aspects of multi-interest modeling can be categorized into user-oriented and item-oriented dimensions based on the source of external information, as illustrated in Figure~\ref{fig:interest_aspect}. More concretely, user-oriented multi-interest modeling aims to extract multiple interests from the external user information, such as temporal factors (\eg short-term and long-term interests, periodic patterns), spatial information, interaction behaviors (\eg clicks, purchases, add to cart or favorites), and social groups. In contrast, item-oriented methods utilize item-associated attributes, including brands, categories, domain features, and reviews, for multi-interest modeling.
Overall, no matter whether item-oriented or user-oriented approaches, the external side information is required to reveal the degree of user preference for each specific aspect. Therefore, we identify these approaches as explicit multi-interest modeling. 

\subsubsection{User-Oriented Multi-Interest Modeling}

User-oriented multi-interest modeling utilizes user profiles, behaviors, and other side information for multi-interest extraction and modeling. The detailed introduction of this category is below.

\paratitle{Spatial Information.} It is noteworthy that users' travels, accommodations, and activity behaviors vary with location changes. For instance, when visiting an international modern metropolis, a user may prioritize shopping malls and museums. Conversely, in a historical and cultural city, the user may prefer natural scenery and historical monuments. While shopping malls and historical monuments both represent user preferences, the variation in location highlights distinct interests.
Consequently, spatial and location information is critically important for multi-interest modeling, particularly for online travel platforms and lifestyle entertainment applications~\cite{tao2022sminet}.

\paratitle{Temporal and Periodic Information.} Users' interest will shift over time. Numerous works aim to model the interest evolution process, and the user's long-term and short-term  preferences~\cite{chen2021exploring, deng2022leveraging, li2024disentangle}. Additionally, users' interests and behavior patterns often exhibit periodic trends. For example, a sports enthusiast may prefer surging or water sports during the summer, while skiing and ice sports become more appealing in the winter. Consequently, incorporating temporal and periodic information is essential for effective multi-interest modeling. 

\paratitle{Social Group Information.} Users' behaviors and potential interests often exhibit the herding phenomenon~\cite{xie2024analytical}, \ie individuals with similar social statuses tend to make similar decisions. For example, highly educated and well-off ladies may prefer products such as perfume, fashion, and cosmetics. By considering the preference similarities among users within the same group, we can utilize group information to enrich the multi-interest profiles of users~\cite{tao2022sminet, jiang2020aspect}.

\paratitle{User Behaviors.} Users' multi-types of behaviors (\eg search, click, comments, add to cart, add to favorite, and purchase) widely exist in most real-world e-commerce recommendation systems, which can reveal users’ latent interests and preferred degrees to different items~\cite{meng2023coarse, li2024multi, wu2024multi}. 
Therefore, investigating users' multi-interests across multiple behavioral dimensions and patterns in a dedicated manner is beneficial to capturing users' fine-grained interests and enhancing both explainability and recommendation performance.

\subsubsection{Item-Oriented Multi-Interest Modeling}

On the item side, multi-interest can also be modeled from the following four types of side information.

\paratitle{Attributes Information.} Similar to the multi-faceted preferences of users, items also possess multiple attributes, such as categories, tags, brands, and corresponding knowledge entities~\cite{zheng2024facetcrs, wang2021popularity, zheng2024mitigating}. Users may exhibit varying preferences for different attributes. For instance, an individual who frequents the gym may prioritize the nutritional content of foods, while fashion elite may prefer clothing from designer brands over mass-market brands. Consequently, this associated meta-information can serve as external knowledge to fine-grained multi-interest modeling and recommendation.

\paratitle{Reviews Information.} 
Item reviews contain enriched external information that provides a comprehensive and accurate item description. Since these reviews are provided by consumers, they are often more persuasive than the descriptions from merchants while simultaneously revealing the user's true preferences.
Extracting high-quality and useful information from reviews is vital for both item and user modeling, as it helps to deeply understand user interests and item characteristics, thereby enhancing multi-interest recommendation~\cite{zheng2024facetcrs, zheng2024hycorec, zheng2024diversity}.

\paratitle{Multi-Modality Information.} Conventional methods often identify the item by a unique ID for recommendation, neglecting the rich information provided by multi-modal data, including numerical and categorical attributes, pictures, videos, and text from the item's brief introduction or user comments.
Owing to the remarkable success of large models in both computer vision (CV) and natural language processing (NLP), multi-modality recommendation has recently become a cutting-edge area of research~\cite{zheng2024hycorec, wang2023missrec}. Consequently, a significant number of works are now dedicated to exploring the user's multi-interests encapsulated in multi-modality information for recommendation~\cite{huhorae, tang2023towards}.

\paratitle{Domain Information.} In practice, users often engage with multiple domains and platforms for different intentions. As exemplified by the book and movie recommendations, a user who enjoys reading literature is more likely to appreciate art movies. Additionally, the preferred genre of books can also influence preferences in the movie domain. Therefore, effectively leveraging item domain information and facilitating knowledge transfer across domains can enhance multi-interest recommendation and also alleviate the cold start problem~\cite{sun2023remit, liu2023joint, jiang2022adaptive, huai2023m2gnn, huhorae, tang2023towards}.

\begin{figure}[]
\centering
\includegraphics[width=0.85\columnwidth]{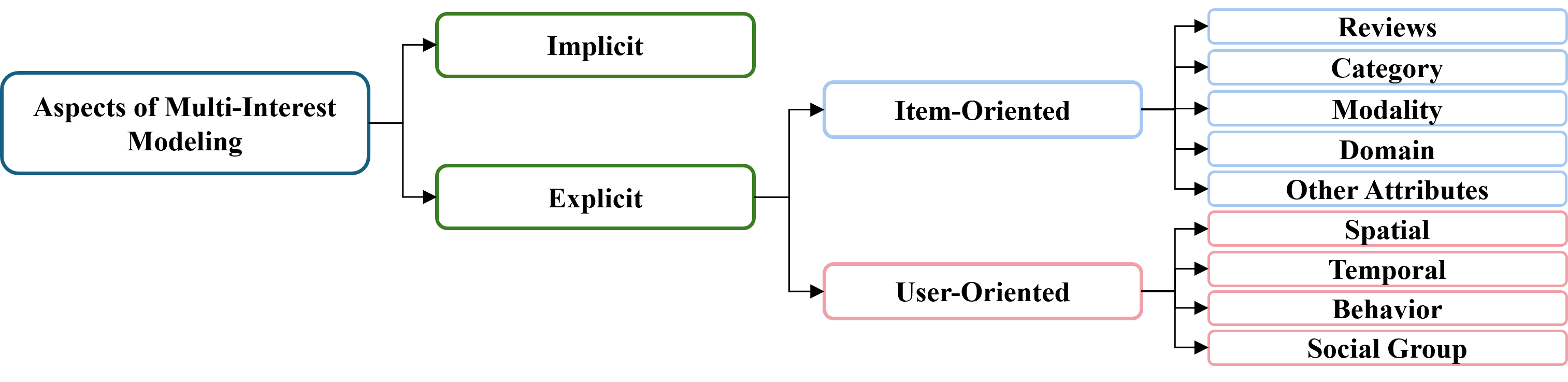}
\caption{The categorization of multi-interest modeling aspects. It can be broadly divided into two categories (implicit multi-interest modeling vs. explicit multi-interest modeling) based on whether external information is used in multi-interest modeling. Further, the explicit multi-interest modeling can be sub-classified as either item-oriented or user-oriented, differentiated by the information associated with items or users.}
\label{fig:category_interest_aspect}
\end{figure}

\subsubsection{Implicit Multi-Interest Modeling} 

Distinct from explicit multi-interest modeling, implicit multi-interest modeling strategies learn the multi-interest representations of users or items from their historical interactions straightforwardly, without external side information incorporation.
As no user or item auxiliary information requirements, implicit multi-interest modeling is a simple yet effective approach and has become a predominant solution in multi-interest recommendation, such as DMIN~\cite{xiao2020deep}, ComiRec~\cite{cen2020controllable}, TimiRec~\cite{wang2022target}.
Specifically, these methods first predefine a fixed number of multiple interests.
Then, sophisticated modules, \eg dynamic routing~\cite{wang2022target, cen2020controllable, du2024disentangled}, or the attention mechanism and its variants~\cite{cen2020controllable, li2022improving, li2020mrif}, are applied to model the multi-interest representations as a vector set.
Therefore, the final recommendations can be obtained based on Eq~\eqref{eq:score_rcf} or Eq~\eqref{eq:score_rpf}.
In conclusion, the aspects of multi-interest modeling can be categorized in Figure~\ref{fig:category_interest_aspect}.

\begin{table}[]
\caption{Classification of representative methods in multi-interest recommendation. CTR means click-through rate prediction.}
\tabcolsep=0.07cm
\small
\begin{tabular}{llll}
\hline\hline
\multicolumn{1}{l}{\textbf{Tasks}}             & \multicolumn{1}{l}{\textbf{Multi-Interest Aspects}}                  & \multicolumn{1}{l}{\textbf{Research Concerns}}                            & \multicolumn{1}{l}{\textbf{Methods}}                                                                                                                                                                                                                                                                      \\ \hline\hline
\multirow{10}{*}{Sequential}    & Behaviors                                                            & Diversity, Multi-behaviors Information Modeling    & \cite{qin2022multi, wu2024multi, li2024multi}                                                                                                                                                                            \\
                                               & Time Periodic \& Temporal Range Information                 & Periodic and Temporal Range Information Modeling   & \cite{chen2021exploring, li2020mrif, huhorae}                                                                                                                                                                                                                        \\
                                               & Multi-modality Information                                           & Multi-modality Information Modeling                & \cite{wang2023missrec}                                                                                                                                                                                                                                                                     \\
                                               & Category                                                             & Category Information Modeling                      & \cite{li2023multi}                                                                                                                                                                                                                                                                             \\
                                               & Attributes                                                           & Interest Representation Collapse Issue                                & \cite{liu2024attribute}                                                                                                                                                                                                                                                                     \\ \cline{2-4} 
                                               & \multirow{5}{*}{Implicit}                                            & Multi Representations Modeling for User and Item                      & \cite{li2019multi, cen2020controllable, wang2022target}                                                                                                                                                \\
                                               &                                                                      & Interest Representation Collapse Issue                                & \cite{du2024disentangled, xie2023rethinking, li2022improving, zhang2022re4}                                                                                          \\
                                               &                                                                      & Tail Item Recommendation                                              & \cite{liu2023co}                                                                                                                                                                                                                                                                             \\
                                               &                                                                      & Graph-based Multi-level Representations Modeling                      & \cite{tian2022multi}                                                                                                                                                                                                                                                                          \\
                                               &                                                                      & Diversity                                                             & \cite{chen2021multi}                                                                                                                                                                                                                                                                        \\ \hline
\multirow{5}{*}{CTR}            & Category                                                             & Item Multi and Hierarchical Representations Modeling                    & \cite{li2022miner, pei2024rimirec}                                                                                                                                                                                                                      \\
                                               & User Group, Time Scale                                               & Multi-time Scale Modeling                          & \cite{jiang2020aspect}                                                                                                                                                                                                                                                                        \\
                                               & Behaviors                                                            & Multi-behaviors Information Modeling               & \cite{meng2023coarse}                                                                                                                                                                                                                                                                         \\
                                               & User and Item attributes                                             & Attributes Modeling                                & \cite{chai2022user}                                                                                                                                                                                                                                                                          \\
                                               & Implicit                                                             & Popularity in Multi-interest Modeling                                 & \cite{wang2021popularity}                                                                                                                                                                                                                                                                     \\ \hline
\multirow{2}{*}{Session-based}  & Spatial \& Temporal Information, User Group                          & Cold Start                                                            & \cite{tao2022sminet}                                                                                                                                                                                                                                                                        \\
                                               & Implicit                                                             & Graph-based Multi Representations Modeling                             & \cite{wang2023modeling}                                                                                                                                                                                                                                                                     \\ \hline
\multirow{2}{*}{Conversation} & Keyword-, Entity-, Context-, Review-facets & Diversity                                                             & \cite{zheng2024facetcrs, zheng2024hycorec, zheng2024mitigating, zheng2024diversity}                                                                                         \\
                                               & Attributes                                                           & Fuzzy Feedback and Attributes Information Modeling & \cite{shen2024multi, zhang2022multiple}                                                                                                                                                                                                                  \\ \hline
Cross-domain                    & Domains and Tag                                                      & Cold Start, Multi-Domains Information Modeling     & \cite{sun2023remit, jiang2022adaptive, zang2023contrastive, liu2023joint, zhang2022diverse, huai2023m2gnn} \\ \hline\hline
\end{tabular}
\label{tab:classfication}
\end{table}

\subsection{Classification of Multi-Interest Recommendation}
\label{sec:classification}

Reviewing existing works, the representative multi-interest recommendation methods can be categorized in Table~\ref{tab:classfication} based on the multi-interest recommendation tasks, multi-interest modeling aspects (ref. Figure~\ref{fig:category_interest_aspect} for detail), and the specific concerns aiming to address. Beyond the fine-grained user and item representation modeling and diversity recommendation enhancement introduced in Subsection~\ref{sec:why}, the concerns of these works also include:

\paratitle{Alleviating Multi-Interest Representation Collapse Issue.} As multi-interest recommendation requires modeling multiple representations of users or items, these representations are prone to encountering the collapse issue, \ie the multi-interest representations collapse into similar ones, which hinders the model from capturing the discriminative information in real-world applications for recommendation. To overcome this issue, a prevalent solution involves incorporating a regulation term (\eg contrastive learning~\cite{du2024disentangled, zhang2022re4}) into the loss functions to push away the interest representations from each other, thereby facilitating discriminative representation learning.
    
\paratitle{Cold Start and Fairness Recommendation.} The challenges associated with cold start and fairness recommendation stem from data sparsity, a prevalent issue in recommendation tasks. In the context of multi-interest recommendation, multiple works~\cite{tao2022sminet, zhang2022diverse} endeavor to incorporate auxiliary information (\eg spatial and temporal information, domain information, user social group, user and item tags) from the user side and the item side to mitigate this problem.

\paratitle{Others.} As the popularity of items affects the user click probability inherently, in~\cite{wang2021popularity}, the authors consider the popularity to model the degree of users’ intention to popular items for the multi-interest click-through rate prediction. 
In~\cite{shen2024multi}, the authors claim that a user is more likely to provide a somewhat fuzzy response like \textit{I don't know} compared to a clear answer, \eg \textit{yes or no}, in a practical recommendation system. Therefore, it is critically important to focus on the fuzzy feedback of users for potential multi-interest intentions modeling and recommendation.

\section{Methodology for Multi-Interest Recommendation}
\label{sec:method}

\subsection{Definition of Multi-Interest Recommendation}
\label{sec:def_tasks}

\begin{figure}[]
\centering
\includegraphics[width=\columnwidth]{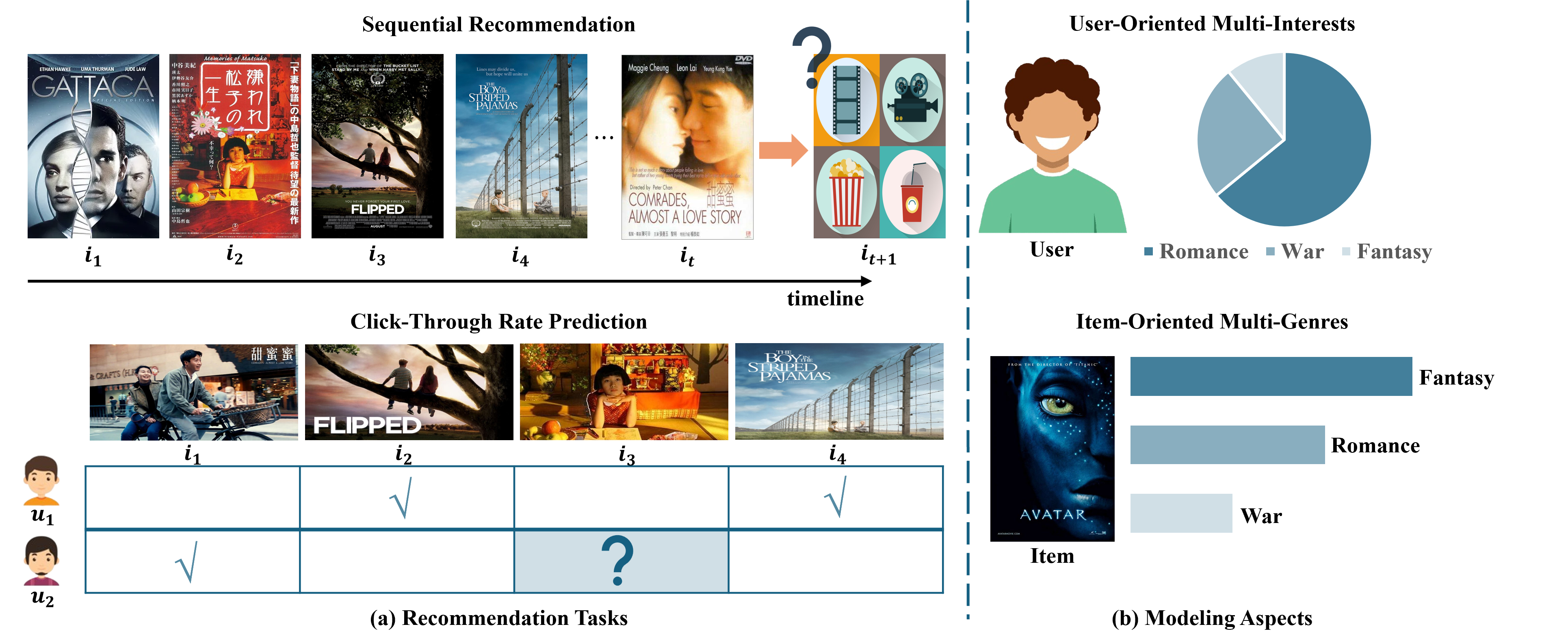}
\caption{Toy example of multi-interest modeling in recommendation tasks. The left diagram depicts two typical recommendation tasks: sequential recommendation and click-through rate prediction. The right part illustrates the user's multi-interest and the item's multi-themes.}
\label{fig:task}
\end{figure}

Reviewing existing works, in general, two primary tasks have garnered extensive exploration in multi-interest recommendation: (1) click-through rate (CTR) prediction, and (2) sequential recommendation, as illustrated in Figure~\ref{fig:task}. We formalized these tasks as below. 

\paratitle{Click-through Rate Prediction.} Given a set of item $\mathcal{I}=\{i_1,i_2,...i_{|\mathcal{I}|}\}$ and a set of user $\mathcal{U}=\{u_1,u_2,...u_{|\mathcal{U}|}\}$, where $|\mathcal{U}|$ and $\mathcal{|I|}$ are the number of users in set $\mathcal{U}$ and items in set $\mathcal{I}$, respectively. Denote the interacted user-item records as a matrix $\mathbf{R}=[r_{i,j}]_{|\mathcal{U}|\times|\mathcal{I}|}$. Specifically, we define the $r_{i,j}=1$ if the user $u_i$ interacted with item $i_j$ in the history records; otherwise $r_{i,j}=0$. Consequently, given any user $u\in \mathcal{U}$, the click-through rate prediction aims to estimate the probability that a user $u$ will click on a given item $i\in \mathcal{I}$  when it is displayed to them.
This process can be formalized as:
\begin{equation}
    P(i|u, \mathbf{R})=f_{\theta}(u,i,\mathbf{R})\label{eq:ctr}
\end{equation}
where $f_{\theta}(\cdot)$ denotes the recommendation model.

\paratitle{Sequential Recommendation.} Distinct from click-through rate prediction, sequential recommendation endeavors to model the user's preference evolution and temporal patterns (\eg long-term and short-term preference) for next item prediction. Therefore, given any user $u$, whose interacted items are organized as a sequence $\mathcal{S}_u$ chronologically, base on the times-tamp $t$, \ie $\mathcal{S}_u=\{i_{1}, i_{2}, ..., i_{t}\}$. The sequential recommendation aims to predict the items that the users would probably like to engage with at $t+1$ time step, which can be formalized as:
\begin{equation}
    P(i_{t+1}|i_1,i_2,...i_t)=f_\theta(i_1,i_2,...,i_t)\label{eq:sq}
\end{equation}

\paratitle{Multi-Interest Recommendation.} In general, under the representation learning (also know as deep learning) framework, both above two tasks require learning user representation $\mathbf{h}_u\in\mathbb{R}^{1\times d}$ and item representation $\mathbf{x}_i\in\mathbb{R}^{1\times d}$, respectively, where $d$ is the dimension of the representation vector. Therefore, the inner product between $\mathbf{h}_u$ and $\mathbf{x}_i$ is calculated as the predicted score $\hat{y}_{u_i}$ of the item $i$ for recommendation:
\begin{equation}
    \hat{y}_{u_i}=\mathbf{h}_u\mathbf{x}_i^\top\label{eq:score}
\end{equation}

As shown in Figure~\ref{fig:task}, considering the inherent uncertainty of users' intentions, the complexity of behavior patterns, and the ambiguous themes of items, it is insufficient to define the user or item representation as a single vector as in conventional methods~\cite{li2019multi, cen2020controllable}.  
Therefore, multi-interest recommendation aims to learn users' multi-interest~\cite{li2022miner, li2019multi, cen2020controllable, li2023exploiting}, items' multi-aspect~\cite{zheng2024facetcrs, wang2021popularity, zheng2024hycorec}, or both~\cite{tao2022sminet, jiang2022adaptive, wang2022target}, through multiple vector representations for an accurate recommendation. 
In this way, the user or item representations in Eq~\eqref{eq:score} are expanded from single vector into a list of vectors, \ie $\mathbf{H}_u=[\mathbf{h}_{u}^1, \mathbf{h}_{u}^2, ..., \mathbf{h}_{u}^K]$ or $\mathbf{X}_i=[\mathbf{x}_{i}^1, \mathbf{x}_{i}^2, ..., \mathbf{x}_{i}^K]$, where $K$ denotes the number of user interests or item aspects. 
Accordingly, the aggregation operation is required to aggregate the multiple interest representations for the final recommendation. Reviewing existing works, two typical aggregation paradigms in multi-interest recommendation can be defined, \ie multi-interest representation aggregation and multi-interest recommendation aggregation (ref. Section~\ref{sec:agg} for detail). Specifically, 

\begin{itemize}
    \item \textbf{Representation Aggregation for Multi-Interest Recommendation.} Although sophisticated methods are dedicated to multi-interest representation modeling, these representations are required to fuse as a single vector for recommendation prediction~\cite{wang2021popularity, wang2022target}. This process can be formalized below:
    \begin{equation}
         \hat{y}_{u_i}=\phi_u(\mathbf{H}_u)\phi_i(\mathbf{X}_i)^\top\label{eq:score_rpf}
    \end{equation}
    where $\phi_u(\cdot)$ and $\phi_i(\cdot)$ are aggregation functions, which can be pooling, concat, attention mechanism, or neural networks. 
    \item \textbf{Recommendation Aggregation for Multi-Interest Recommendation.} In contrast to representation aggregation, recommendation aggregation requires attaining the recommendations for each interest based on Eq~\eqref{eq:score} first, and then combining these scores through specific strategies, such as the max operation~\cite{xie2023rethinking, zhang2022re4, liu2024attribute}, to obtain the final recommendation results, which can be formalized as below:
    \begin{equation}
        \hat{y}_{u_i}=\phi(\mathbf{H}_u\mathbf{X}_i^\top)\label{eq:score_rcf}
    \end{equation}
    where $\phi(\cdot)$ represents the strategy used to select the items that the user most likes from each interest predicted result, thereby forming the final recommendation.
\end{itemize}

\begin{figure}[]
\centering
\includegraphics[width=\columnwidth]{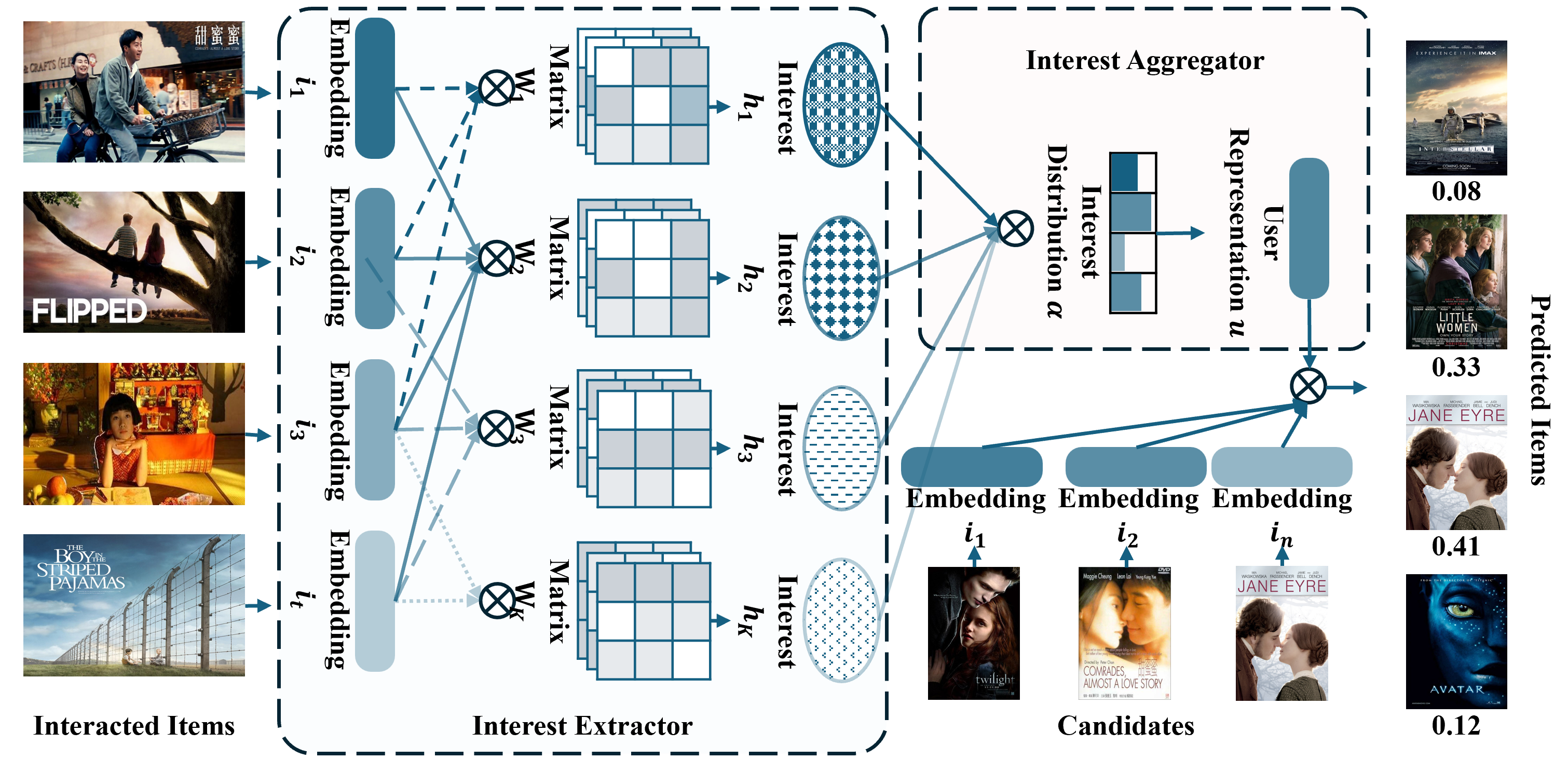}
\caption{The framework of multi-interest modeling and recommendation. It includes two main components: an interest extractor and an interest aggregator, circled by the dotted line boxes respectively.}
\label{fig:framework}
\end{figure}


Consequently, as shown in Figure~\ref{fig:framework}, the overall framework of multi-interest recommendation consists of two main components: the \textit{interest extractor} and the \textit{interest aggregator}. Specifically, the interest extractor is designed to learn the user's multiple interest representations from the interacted items and associated side information. Subsequently, these multiple interest representations are fed into the interest aggregator, which either fuses the representations or combines the corresponding recommended items derived from each interest representation to obtain the final results. In the following subsections, we provide a comprehensive introduction to the technical details of each component.

\subsection{Multi-Interest Extractor}
\label{sec:ext}

\begin{figure}[h!]
\centering
\includegraphics[width=\columnwidth]{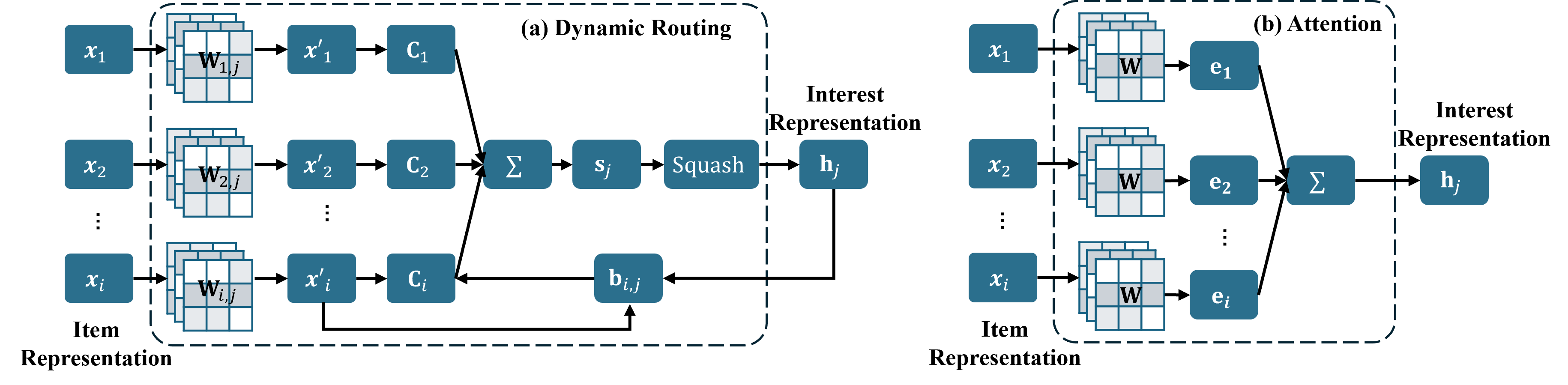}
\caption{The architectures of multi-interest extraction (dynamic routing vs. attention). Given the interacted item representations $\{\textbf{x}_1,\textbf{x}_2, ..., \textbf{x}_i\}$, by defining different learnable matrices $\mathbf{W}$ and initializing multi-interest vectors $\mathbf{e}$, dynamic routing and attention can be employed to extract multiple interest representations. 
Here we take $\mathbf{h}_j$ as an example.}
\label{fig:extractor}
\end{figure}

With respect to the multi-interest modeling, reviewing existing literature, there are several representative methods, including dynamic routing, attention and its variants, iterative attention, and non-linear transformation. Among these, dynamic routing with capsule neural networks and attention mechanisms (along with their variants) are the most widely adopted approaches for multi-interest extraction. The architectures of these two methods are illustrated in Figure~\ref{fig:extractor}. A detailed introduction to these methods is provided below.

\paratitle{Dynamic Routing.}
As a novel neural network architecture, CapsNet~\cite{sabour2017dynamic} specializes in hierarchical structure learning.
In general, CapsNet defines a series of capsules, which are collections of neurons that can be individually activated based on various properties of an object, such as position, size, and hue. Consequently, different objects can elicit distinct activity vectors from the capsule. Note that the length of the activity vector represents the probability of the presence of the properties in the given object, while the orientation of the vector indicates the strength of these properties within the capsule.
Dynamic routing is the iterative algorithm proposed in CapsNet for capsule representation learning.
In multi-interest modeling, we define the interest capsule as a vector set $\mathbf{H}=[\mathbf{h}_1, \mathbf{h}_2, ..., \mathbf{h}_K]$, which is updated by a non-linear squashing function~\cite{sabour2017dynamic}, as below: 
\begin{equation}
    \mathbf{h}_j=\textrm{squash}(\mathbf{s}_j)=\frac{||\mathbf{s}_j||^2}{1+||\mathbf{s}_j||^2}\frac{\mathbf{s}_j}{||\mathbf{s}_j||}
\end{equation}
where $||\mathbf{s}_j||$ denotes the Euclidean norm (also known as $L^2$ norm), which measures the magnitude (\ie length) of the vector $\mathbf{s}_j$. Thus, $\textrm{squash}(\cdot)$ function can be recognized as an activation function, which is designed to suppress short vectors, driving their values toward zero, while compressing long vectors toward approaching one. $\mathbf{s}_j$ is calculated as,  
\begin{equation}
    \begin{split}
        &\mathbf{x}_i'=\mathbf{W}_{ij}\mathbf{x}_i\\
        &\mathbf{s}_j=\sum_{i=1}^tc_{ij}\mathbf{x}_i'\\
        &c_{ij}=\textrm{softmax}(b_{ij})=\frac{\textrm{exp}(b_{ij})}{\sum_{j=1}^K\textrm{exp}(b_{ij})}\\
        &b_{ij}=b_{ij}+\mathbf{h}_j\mathbf{x}_i'^\top\label{eq:dynamic_routing}
    \end{split}
\end{equation}
where $\mathbf{W}_{ij}$ is learnable transformation weight matrices. $c_{ij}$ is the coupling coefficient between item $i$ and all the capsules, generated by the iterative dynamic routing. $b_{ij}$ denotes the log prior probability that item $i$ should be coupled to capsule $j$, and it is initialized to zero. The whole iterative process of the dynamic routing is illustrated in Algorithm~\ref{alg:dyrouting}.

\RestyleAlgo{ruled}
\begin{algorithm}[t]
	\renewcommand{\algorithmicrequire}{\textbf{Adapter-fused Encoder Fine-tuning:}}
	\renewcommand{\algorithmicensure}{\textbf{Output:}}
    \renewcommand{\algorithmicensure}{\textbf{Process:}}
	\caption{Dynamic Routing}
	\label{alg:encoder}
	\begin{algorithmic}[1]
            \STATE \textbf{Input:}
            \STATE \quad Iteration Rounds: $R$\;
            \STATE \quad Item Representation: $[\mathbf{x}_1, \mathbf{x}_2,...,\mathbf{x}_t]$\; 
            \STATE \quad Initialized Routing Logit $b_{ij}$: $b_{ij}=0$\;
            \STATE \textbf{Output:}
            \STATE \quad Multi-interest Representations: $[\mathbf{h}_1,\mathbf{h}_2,...,\mathbf{h}_K]$;
		\STATE \textbf{Process:}
        \FOR{$r=1,2,..,R$}
            \STATE for each low-level capsule (item) $i$, calculate $c_{ij}$: $c_{ij}=\textrm{softmax}(b_{ij})=\frac{\textrm{exp}(b_{ij})}{\sum_{j=1}^K\textrm{exp}(b_{ij})}$; 
            \STATE for each high-level capsule (interest) $j$, calculate $\mathbf{s}_j$: $\mathbf{s}_j=\sum_{i=1}^tc_{ij}\mathbf{W}_{ij}\mathbf{x}_i$;  
            \STATE  for each high-level capsule $j$, calculate $\mathbf{h}_j$: $\mathbf{h}_j=\textrm{squash}(\mathbf{s}_j)=\frac{||\mathbf{s}_j||^2}{1+||\mathbf{s}_j||^2}\frac{\mathbf{s}_j}{||\mathbf{s}_j||}$;
            \STATE update $b_{ij}$: $b_{ij}=b_{ij}+\mathbf{h}_j(\mathbf{Wx}_i)^\top$;
            \ENDFOR
        \RETURN Multi-interest Representations: $[\mathbf{h}_1,\mathbf{h}_2,...,\mathbf{h}_K]$; 
    \end{algorithmic}
    \label{alg:dyrouting}
\end{algorithm}

\paratitle{Attention-Aware Multi-Interest Extractor.} Instead of using the dynamic routing, another prevalent solution is to utilize the attention mechanism for multi-interest modeling. Suppose $\mathbf{E}=[\mathbf{e}_1,\mathbf{e}_2,..., \mathbf{e}_K]$ are the initialized multi-interest embeddings, which are learnable vectors.  
Attention-aware interest extractor dedicated to learning multiple diverse interest representations $\mathbf{H}=[\mathbf{h}_1, \mathbf{h}_2, ... ,\mathbf{h}_K]$ from user's interacted item representations $[\mathbf{x}_1, \mathbf{x}_2,..,\mathbf{x}_t]$. Specifically, this process can be formalized as follows,
\begin{equation}
    \begin{split}
        &\mathbf{h}_j=\sum_{i=1}^tw_i^j\mathbf{x}_i,\quad\quad j=1,2,...,K\\
        &w_i^j=\textrm{softmax}\left(\mathbf{e}_j\sigma(\mathbf{W}_j\textbf{x}_i+\mathbf{b}_j)^\top \right)=\frac{\textrm{exp}\left(\mathbf{e}_j\sigma(\mathbf{W}_j\mathbf{x}_i+\mathbf{b}_j)^\top/\tau\right)}{\sum_{i=1}^t\textrm{exp}\left(\mathbf{e}_j\sigma(\mathbf{W}_j\mathbf{x}_i+\mathbf{b}_j)^\top/\tau\right)}\label{eq:softatt_ext}
    \end{split}
\end{equation}
where $\mathbf{x}_i$ is the representation of item $i$, $\mathbf{W}_j$ and $\mathbf{b}_j$ are learnable parameters associated with the $j$-th interest. Variable $\tau$ is a temperature ratio that controls the sharpness of the distributions, thereby regulating the allocation of attention to the items for the interest representation generation. Considering the extensive computation costs induced by multiple learnable matrices, most works only define a uniform $\mathbf{W}$ and $\mathbf{b}$ for all the interest representations. $\sigma(\cdot)$ is the activation function, \eg tanh in~\cite{li2022miner, wang2021popularity}. Additionally, the activation function, learnable parameters $\mathbf{W}$ and $\mathbf{b}$ can be omitted in some works~\cite{li2022miner, wang2022target}.  

\paratitle{The Variation of Attention-Aware Multi-Interest Extractor.}
In addition to the temperature ratio, in~\cite{huai2023m2gnn} the authors compute the value of $\mathbf{e}_j^{\top}\sigma(\mathbf{W}_j\textbf{x}_i+\mathbf{b}_j)$ raised to a given power, \ie $\frac{\textrm{exp}\left(\textrm{pow}\left(\mathbf{e}_j\sigma(\mathbf{W}_j\textbf{x}_i+\mathbf{b}_j)^\top, \gamma \right) \right)}{\sum_{i=1}^t\textrm{exp}\left(\textrm{pow}\left(\mathbf{e}_j^\top\sigma(\mathbf{W}_j\textbf{x}_i+\mathbf{b}_j)^\top, \gamma \right) \right)}$, to control the shape of the generated distribution, where $\gamma$ modulates the sharpness of the distribution. 
As $\gamma$ increases, more significant items will receive greater attention, and vice versa, \ie when $\gamma=0$, the softmax function degenerates into average pooling operation, while $\gamma \rightarrow \infty$, the function is transformed into hard attention, which selects the most relevant one as the output.
Rather than computing the inner product between the item and interest representations (\ie $\mathbf{e}^\top\mathbf{x}$), in~\cite{wang2022target}, the authors employ the cosine similarity, \ie $\frac{\mathbf{e} \cdot \mathbf{x}}{||\mathbf{e}|| \, ||\mathbf{x}||}$ for distribution generation. 
Furthermore, in~\cite{wang2023modeling}, the authors apply Gumbel-softmax with hard attention as an alternative to the vanilla attention mechanism. That is, a Gumbel noise $\epsilon$, drawn from the Gumbel distribution with a mean of $0$ and a standard deviation of $1$, is added for attention calculation, \ie $\textrm{Gumbel-softmax}\left(\textrm{log}(\textrm{e}_j\mathbf{x}_i^\top)+\epsilon\right)=\frac{\textrm{exp}\left(\left(\textrm{log}(\textrm{e}_j\mathbf{x}_i^\top)+\epsilon\right)/\tau\right)}{\sum_{i=1}^t \textrm{exp}\left(\left(\textrm{log}(\textrm{e}_j\mathbf{x}_i^\top)+\epsilon\right)/\tau\right)}$.

\paratitle{Non-linear Transformation Multi-Interest Extractor.} In contrast to the aforementioned methods that derive a user's multiple interests from the representation of each interacted item, this approach acquires multi-interest representations based on the user representation with a non-linear transformation operation straightforwardly. Specifically,
\begin{equation}
    \mathbf{h}_j=\textrm{LeakyReLU}\left(\mathbf{uW}_j+\mathbf{b}_j \right),\quad j=1,2,...,K
\end{equation}
where $\mathbf{u}$ is the user representation modeled by a neural network. $\mathbf{W}_j, \mathbf{b}_j$ are the learnable transformation matrix and bias matrix corresponding to the interest $j$.

\paratitle{Others.} Inspired by the dynamic routing in multi-interest modeling, in~\cite{zhang2022multiple, shen2024multi}, the authors choose to design the iterative update process for attention calculation. Apart from dynamic routing and attention mechanisms, some sophisticated modules are also designed to harness different side information for multi-interest recommendation. 
For instance, in~\cite{zheng2024facetcrs, sun2023remit, zheng2024hycorec}, hypergraph neural networks and knowledge graphs are employed for item attributes and knowledge modeling, which will be further considered for multi-interest modeling. 
In~\cite{tao2022sminet}, the authors are dedicated to extracting multi-interest representations from spatial-temporal information with a variant attention neural network. 
Since cross-domain information from user groups and items can enhance the performance in the target domain, in~\cite{jiang2022adaptive}, shared networks and domain-specific networks are proposed respectively to model the commonalities and diversities across different domains. After that, a linear projection with concatenation operations is applied for multi-domain multi-interest modeling. 
In~\cite{wang2023missrec}, a multi-modal pre-training and transfer learning framework is proposed for multi-modality information modeling. Besides, an interest-aware decoder with a lightweight dynamic fusion module is applied for multi-interest extraction and aggregation. By introducing the multi-modality information, the cold start problem can be alleviated effectively. 
Given the complexity and diversity of auxiliary information modeling, it is not the primary focus of this survey, therefore, we do not provide a detailed introduction to this part.

\subsection{Multi-Interest Aggregator}
\label{sec:agg}

\begin{figure}[]
\centering
\includegraphics[width=0.95\columnwidth]{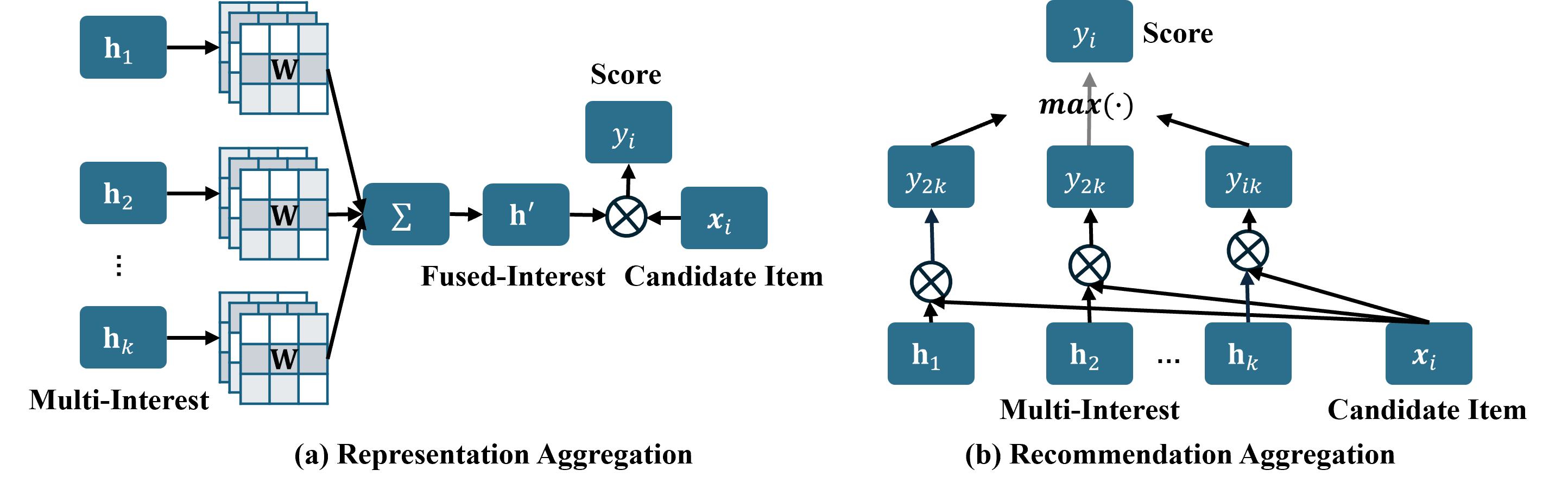}
\caption{The module architectures of multi-interest aggregator (representation aggregation vs. recommendation aggregation).}
\label{fig:aggregator}
\end{figure}

As introduced in Section~\ref{sec:def_tasks}, the methods of multi-interest aggregations can be classified into two categories: representation aggregation and recommendation aggregation. The module architectures of these two techniques are shown in Figure~\ref{fig:aggregator}.  

\subsubsection{Representation Aggregation}
\label{sec:agg_rep}

Representation aggregation aims to fuse the multiple diverse interest representations into a single vector based on dedicated modules or operations. In summary, the representative methods are as follows.

\paratitle{Concat or Pooling.} To obtain the final interest representation $\mathbf{h}'$, the concat or mean/max pooling operations are simple yet effective strategies and widely employed in many works~\cite{tao2022sminet, zheng2024hycorec}, which can be formalized as follows,   
\begin{equation}
    \mathbf{h}'=\textrm{MLP}(\textrm{Concat/Pooling }([\mathbf{h_1},\mathbf{h}_2,...,\mathbf{h}_K]))\label{eq:concat}
\end{equation}
where MLP is a multi-layer perception layer that serves two primary purposes: (1) dimensional transfer and (2) mapping the multiple interest representations from different aspects into a uniform semantic latent space for the concat operation. 

\paratitle{Attention-Aware Aggregation.} 
Note that a user's preference can be decomposed into two distinct components: the primary basic interest and multiple diverse interests. The primary interest is modeled based on the user's historically interacted items, revealing the user's base interest, whereas the multiple diverse interests aim to model the preference to various specific aspects further, such as behaviors of users (\eg click, cart-add, purchase, favor) and multiple themes of movies (\eg romantic, fiction, comedy, war). Consequently, the correlation between the base interest and the multiple interests can be calculated to facilitate the aggregation of multiple interest representations. This process can be formalized as follows:
\begin{equation}
    \begin{split}
        &w_i=\textrm{softmax}(\mathbf{h}_b\mathbf{Wh}_i^\top)=\frac{\textrm{exp}\left(\mathbf{h}_b\mathbf{Wh}_i^\top\right)}{\sum_{i=1}^K\textrm{exp}\left(\mathbf{h}_b\mathbf{Wh}_i^\top\right)}\\
        &\mathbf{h}'=\sum_{i=1}^Kw_i\mathbf{h}_i
    \end{split}
\end{equation}
where $\mathbf{h}_b$ and $\mathbf{h}_i$ are user's basic interest representation and $i$-th multi-interest representation, respectively. $\mathbf{W}$ is the learnable parameters.

\paratitle{Reinforcement Learning for Multi-Interest Selector.} 
In contrast to the aforementioned solutions that consider all interest representations for fusion, some works~\cite{sun2023remit, zhang2022multiple} believe that not all the interests contribute to the current item. Therefore, an alternative approach involves dynamically selecting the most relevant interests through reinforcement learning for representation aggregation.
In~\cite{sun2023remit}, the authors recognize the state $s$ as determined by the user's interest regarding the current items. Therefore, a policy model $\pi_{\theta}(s, a)$, \ie a two-layer feed-forward network with ReLU activation function, is employed to derive the action distribution based on the current state $s$. $a\in\{1,0\}$ denotes the action value, where $1$ indicates the interest is selected for the current item. The recommendation performance is considered as the reward value for policy updates.
Besides, in~\cite{zhang2022multiple, shen2024multi}, the authors adopt dueling Q-network~\cite{wang2016dueling}, which defines the state $s_t$ as the current interest. Accordingly, the action $a_t$ represents the predicted items based on the current interest. Therefore, following the Bellman equation~\cite{bellman1957role}, the maximum expected reward $Q^*(s_t,a_t)$ under the optimal policy $\pi^*$ is given by:
\begin{equation}
    \begin{split}
        & Q^*(s_t,a_t)=\mathbb{R}_{s_{t+1}}\left[r_t+\gamma \max_{a_{t+1}\in\mathcal{A}_{t+1}} Q^*(s_{t+1},a_{t+1}|s_{t},a+t) \right]\\
        &Q(s_t,a_t)=\max_i\left(f_{\theta_V}(\mathbf{h}_i)+f_{\theta_A}(\mathbf{h}_i,a_t) \right)\quad i=1,2,...,K
    \end{split}\label{eq:qvalue}
\end{equation}
where $f_{\theta_V}(\cdot)$ and $f_{\theta_A}(\cdot)$ are two discrete MLP networks. $\gamma$ is a discount factor, which diminishes the reward per timestep.

\subsubsection{Recommendation Aggregation}
\label{sec:agg_rec}

The principle of recommendation aggregation is to attain the items that a user is most likely to prefer from the recommendation pool based on the predicted scores associated with each interest. Typically, there are the following representative strategies:

\paratitle{Mean/Max Pooling.} Let $Y_i=[y_i^1, y_i^2,...,y_i^K]$ as the predicted scores of item $i$ derived from Eq.~\eqref{eq:score}, $K$ is the number of total interests. The mean/max pooling operation takes the average or max value of the $Y_i$ as the final recommendation score of item $i$, \ie
\begin{equation}
    y_i'=\textrm{Mean/Max }(Y_i)=\textrm{Mean/Max }([y_i^1, y_i^2,...,y_i^K])\label{eq:agg_meanmax}
\end{equation}

\paratitle{Attention-Aware Aggregation.} Different from the pooling operation, a more dynamic and adaptive manner is to adjust the weight of each predicted score for the summation, where each weight is defined based on the relevance between the target item representation $\mathbf{x}_i$ and the current interest representation $\mathbf{h}_j$. Consequently, the attention mechanism is formalized with:
\begin{equation}
    \begin{split}
        & y_i'=\sum_{j=1}^Kw_jy_i^j \\
        & w_j= \textrm{softmax}\left(\mathbf{h}_j\sigma(\mathbf{Wx}_i)^\top\right)=\frac{\textrm{exp}\left(\mathbf{h}_j\sigma(\mathbf{Wx}_i)^\top\right)}{\sum_{j=1}^K\textrm{exp}\left(\mathbf{h}_j\sigma(\mathbf{Wx}_i)^\top\right)}\\
    \end{split}
\end{equation}
where $\sigma(\cdot)$ is the activation function, such as GELU in~\cite{li2022miner}. $\mathbf{W}$ are learnable parameters. In~\cite{li2020mrif}, the activation function and learnable parameters are removed for recommendation fusion.

\begin{table}[]
\caption{Comparison of representative multi-interest recommendation methods from the multi-interest extractor and multi-interest aggregator, two perspectives.}
\tabcolsep=0.07cm
\small
\begin{tabular}{llll}
\hline\hline
\textbf{Interest Extractor}      & \multicolumn{2}{l}{\textbf{Interest Aggregator}}                        & \textbf{Representative Methods}              \\ \hline\hline
\multirow{4}{*}{\begin{tabular}[c]{@{}l@{}}Dynamic \\ Routing\end{tabular}} & \multirow{2}{*}{\begin{tabular}[c]{@{}l@{}}Representation \\ Aggregation\end{tabular}}  & Attention             & MIND~\cite{li2019multi} , M2GNN~\cite{huai2023m2gnn}, MDSR~\cite{chen2021multi}                            \\
                                 &                                             & Concat or Mean/Max Pooling & M2GNN~\cite{huai2023m2gnn}, MISD~\cite{li2024multi}                                  \\ \cline{2-4} 
                                 & \multirow{2}{*}{\begin{tabular}[c]{@{}l@{}}Recommendation\\ Aggregation\end{tabular}} & Mean/Max Pooling           & MINER~\cite{li2022miner}, ComiRec~\cite{cen2020controllable}, REMI~\cite{xie2023rethinking}, MGNM~\cite{tian2022multi}, UMI~\cite{chai2022user}              \\
                                 &                                             & Attention             & MINER~\cite{li2022miner}, DisMIR~\cite{du2024disentangled}                                \\ \hline
\multirow{5}{*}{Attention}  & \multirow{3}{*}{\begin{tabular}[c]{@{}l@{}}Representation \\ Aggregation\end{tabular}}  & Concat or Mean/Max Pooling & PENR~\cite{wang2021popularity}, M2GNN~\cite{huai2023m2gnn}                                  \\
                                 &                                             & Attention             & PIPM~\cite{qin2022multi}, M2GNN~\cite{huai2023m2gnn}                                   \\
                                 &                                             & Interest Selector with RL  & REMIT~\cite{sun2023remit}                                        \\ \cline{2-4} 
                                 & \multirow{2}{*}{\begin{tabular}[c]{@{}l@{}}Recommendation \\ Aggregation\end{tabular}}  & Mean/Max Pooling           & MINER~\cite{li2022miner}, TimiRec~\cite{wang2022target}, PIMI~\cite{chen2021exploring}, ComiRec~\cite{cen2020controllable}, CMI~\cite{li2022improving}\\
                                 &                                             & Attention             & MINER~\cite{li2022miner}, MI-GNN~\cite{wang2023modeling}                                \\ \hline
\begin{tabular}[c]{@{}l@{}}Iterative \\ Attention\end{tabular}         & \begin{tabular}[c]{@{}l@{}}Recommendation \\ Aggregation\end{tabular}                    & Interest Selector with RL  & MIMCR~\cite{zhang2022multiple}                                        \\ \hline
\begin{tabular}[c]{@{}l@{}}Non-linear \\ Transformation\end{tabular}        & \begin{tabular}[c]{@{}l@{}}Representation \\ Aggregation\end{tabular}                   & Concat                     & CKML~\cite{meng2023coarse}                                         \\ \hline\hline
\end{tabular}
\label{tab:ext_agg}
\end{table}

In summary, Table~\ref{tab:ext_agg} outlines the representative models featuring the aforementioned dedicated multi-interest extractors and aggregators. It is evident that the attention mechanism is the most frequently applied solution for multi-interest extraction, followed by dynamic routing. Regarding the interest aggregator, using max pooling to fuse the recommendation results derived from each interest representation has become one of the canonical approaches due to its effectiveness and efficiency. Additionally, the attention mechanism is also a popular method for multi-interest representation aggregation.

\subsection{Multi-Interest Representation Regularization for Diversity}
\label{sec:reg}

\begin{figure}[h!]
\centering
\includegraphics[width=0.65\columnwidth]{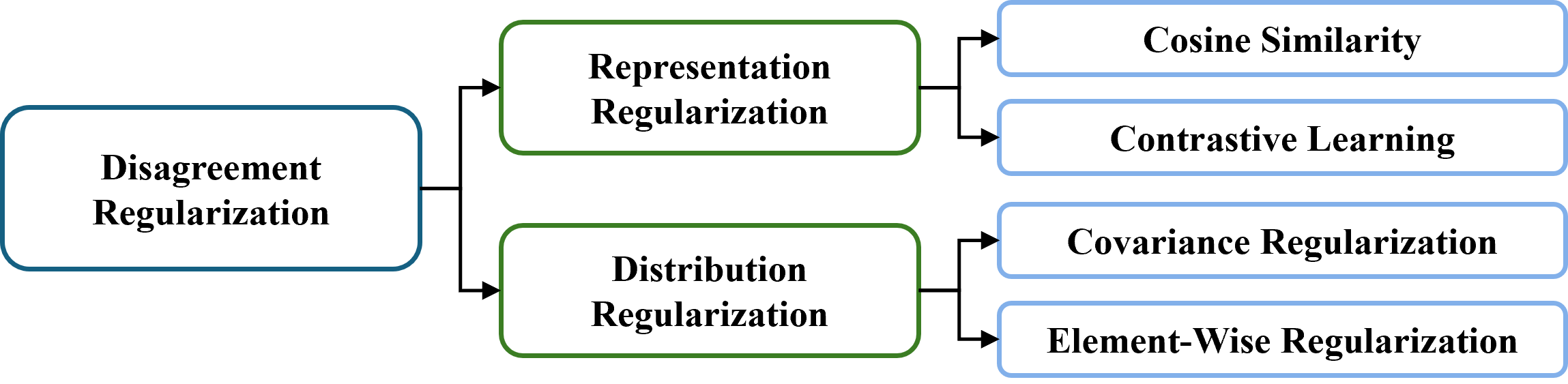}
\caption{The categorization of multi-interest representation regularization for diversity. By increasing the dissimilarity of each representation during optimization, the multiple interest representation is ensured to capture distinct aspects of user behaviors, thereby enhancing the overall performance and effectiveness of the model.}
\label{fig:regularization}
\end{figure}

During the model training, the multi-interest representations derived from the dynamic routing or attention module encounter the potential risk of representation collapse, \ie the model learns a trivial solution where all the interest representations may incline to be indistinguishable from each other and collapse to a narrow point in the hyperspace.
This phenomenon significantly diminishes the representation capability of models and degrades the final performance~\cite{ma2019learning, li2024spectral}. 
To enhance the diversity and distinctiveness of multi-interest representations, a prominent recipe is to incorporate a tailored disagreement regularization term into loss functions. 
Summary of the existing research, the commonly used regularization terms can be categorized into two directions \ie multi-interest representation regularization and multi-interest distribution regularization, as shown in Figure~\ref{fig:regularization}.

\subsubsection{Multi-interest Representation Regularization} 
\begin{figure}[]
\centering
\includegraphics[width=0.95\columnwidth]{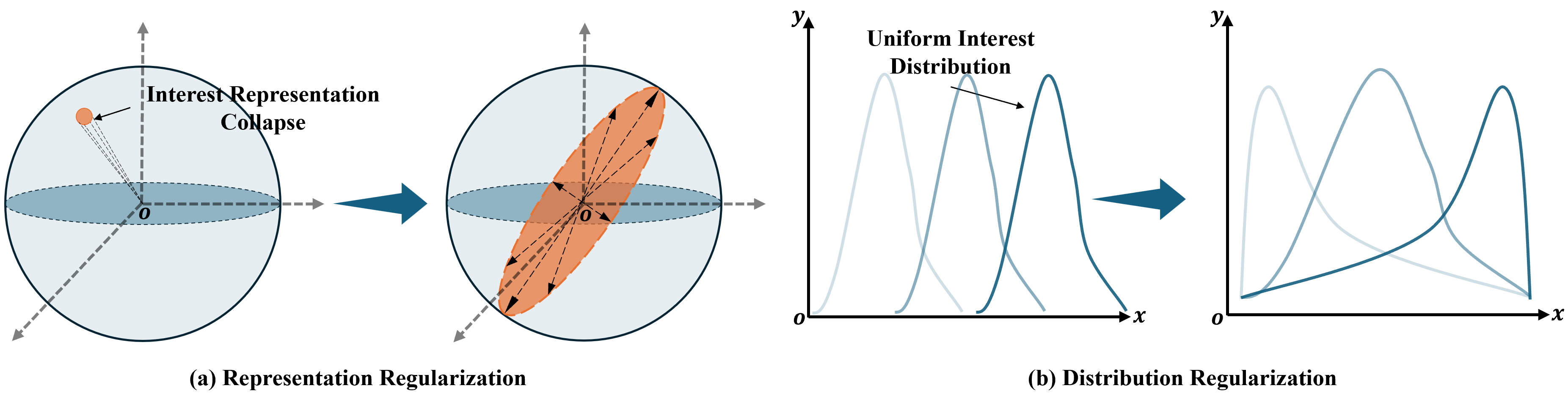}
\caption{The diagram illustrates the process of regularization working on the multi-interest representation hyperspace (left) and the multi-interest representation distribution (right). Representation regularization aims to optimize the interest representations by expanding them from a narrow point to encompass the entire hyperspace. In contrast, distribution regularization endeavors to transform the multi-interest distributions from uniformity to anisotropy.}
\label{fig:regularization_toy}
\end{figure}

Representation regularization aims to enlarge the distance of each interest pair within the semantic hyperspace, alleviating the representation collapse issue, as shown in Figure~\ref{fig:regularization_toy} (left). The representative solutions in this line include:

\paratitle{Cosine Similarity.} In~\cite{li2022miner, chen2021multi, li2022miner}, the authors minimize the cosine similarity across all interest representation pairs to increase the distinctiveness, which can be formalized as below:
\begin{equation}
    \mathcal{L}_{reg}=\frac{1}{K^2}\sum_{i=1}^K\sum_{j=1}^K\frac{\mathbf{h}_i\cdot\mathbf{h}_j}{||\mathbf{h}_i||\,||\mathbf{h}_j||}
\end{equation}
where $\mathbf{h}$ is the interest representation, $K$ is the number of multi-interest. $||\cdot||$ denotes the  Euclidean norm of vectors.
    
\paratitle{Contrastive Learning.}  Inspired by~\cite{wu2018unsupervised, oord2018representation}, contrastive learning is performed by pushing the instances closer to positive samples while distancing the instances from negative samples. Accordingly, InfoNCE loss~\cite{oord2018representation} can be employed as the regularization:
\begin{equation}
    \mathcal{L}_{reg}=-\frac{1}{K}\sum_{i=1}^K\textrm{log}\frac{\exp(\textrm{sim}(\mathbf{h}_i,\mathbf{h}_i^+)/\tau)}{\exp(\textrm{sim}(\mathbf{h}_i,\mathbf{h}_i^+)/\tau)+\sum_{j=1}^n\exp(\textrm{sim}(\mathbf{h}_i,\mathbf{h}_j^-)/\tau)}
\end{equation}
where $\textrm{sim}(\cdot)$ can be the inner product (\ie cosine similarity) or dot product with L2 normalization. The variable $\tau$ is the temperature ratio, which controls the shape of the distribution. In general, the positives $\mathbf{h}^+$ can be yielded by the original representation $\mathbf{h}$ through noise perturbation or dropout operation, while the negatives are randomly sampled for the other interest representations, excluding $\mathbf{h}$. 

\subsubsection{Multi-Interest Distribution Regularization} 
Contrary to the representation regularization, which increases the discriminative ability in the representation hyperspace, distribution regularization aims to mitigate the representation collapse by increasing the variation of each interest distribution, as shown in Figure~\ref{fig:regularization_toy} (right). 

\paratitle{Covariance Regularization}. In probability theory and statistics, the diagonal elements of the covariance represent the variances, thereby indicating the degree of dispersion within the data. Accordingly, we could regularize the item-interest routing matrix $\mathbf{C}\in\mathbb{R}^{t\times K}$ by imposing constraints on its covariance matrix, where $t$ is the number of items and $K$ is the number of interests. This implementation can be formalized as below:
\begin{equation}
\begin{split}
    & \mathrm{Cov}(\mathbf{C}, \mathbf{C})=(\mathbf{C}-\bar{\mathbf{C}})^\top(\mathbf{C}-\bar{\mathbf{C}})\\
    & \mathcal{L}_{reg}=||\textrm{diag}(\mathrm{Cov}(\mathbf{C}, \mathbf{C}))||^2_F
\end{split}
\label{eq:cov}
\end{equation}
where $\textrm{diag}(\cdot)$ represents the diagonal operation. $||\cdot||_F$ denotes the Frobenius norm of matrix. $\bar{\mathbf{C}}$ is the mean value of matrix $\mathbf{C}$ along the first axis. Therefore, $\mathrm{Cov}(\mathbf{C}, \mathbf{C})$ represents the covariance matrix of the routing distribution for each interest.

\paratitle{Element-Wise Regularization.} Distinct from covariance regularization, a more straightforward method is to regulate the differentiation of each interest attention distribution from the matrix element-wise perspective~\cite{li2018multi}, as follows,
\begin{equation}
    \mathcal{L}_{reg}=\frac{1}{K^2}\sum_{i=1}^K\sum_{j=1}^K||\mathbf{W}_i\odot\mathbf{W}_j||\label{eq:att_reg}
\end{equation}
where $\mathbf{W}_i$ and $\mathbf{W}_j$ are the attention distribution matrix of interest $i$ and interest $j$ respectively. $||\cdot||$ denotes the Euclidean norm.

\section{Applications and Public Datasets}
\label{sec:application}


\begin{figure}
\centering
\includegraphics[width=0.95\columnwidth]{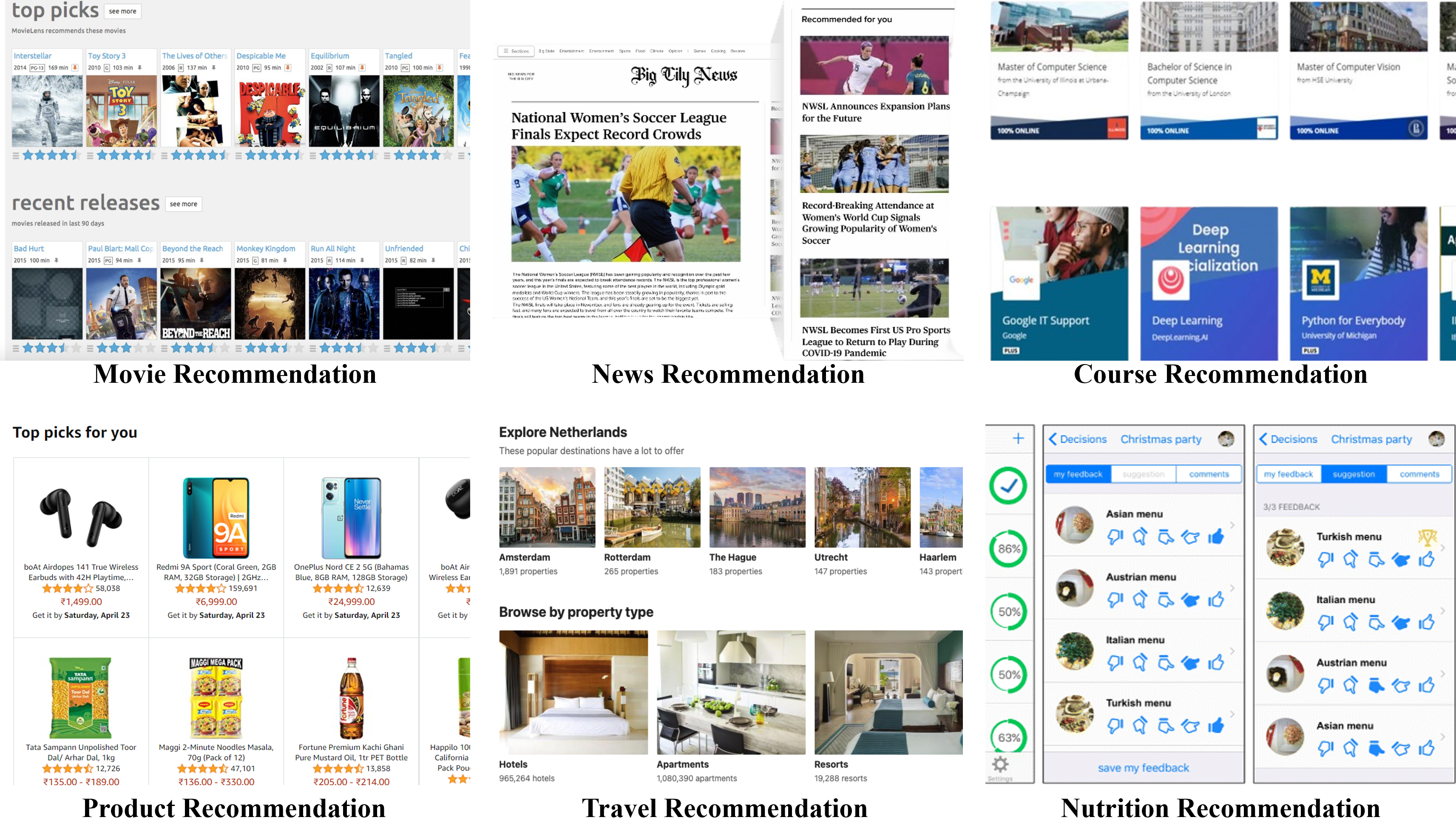}
\caption{Application scenarios of multi-interest recommendation.}
\label{fig:apps}
\end{figure}

Multi-interest modeling has widespread significant attention in recommendation systems and has been extensively adopted across various scenarios, including both common applications and vertical domain tasks, as illustrated in Figure~\ref{fig:apps}. More concretely, vertical domain tasks focus on specialized fields requiring extensive domain knowledge, such as healthcare and education. In contrast, the general task applications endeavor to serve in our daily lives, like online shopping and entertainment. 
We provide a brief overview and analysis of multi-interest recommendation within these varied application contexts below. Most of them have been verified in the real-world online systems.
Additionally, the public real-world datasets employed in these multi-interest recommendation applications are shown in Table~\ref{tab:dataset}.
In the following, we detail the source and composition of each dataset,
respectively.

\begin{table}[]
\caption{Public real-world datasets on different application scenarios.}
\begin{tabular}{ll}
\hline\hline
\textbf{Applications}               & \textbf{Public Datasets}                       \\ \hline\hline
News                       & MIND\tablefootnote{\url{https://msnews.github.io/}}                                  \\
Movies \& Micro Videos     & MovieLens\tablefootnote{\url{https://grouplens.org/datasets/movielens/}}, KuaiShou\tablefootnote{\url{https://www.heywhale.com/home/competition/5ad306e633a98340e004f8d1}}, ReDial\tablefootnote{\url{https://redialdata.github.io/website/}}, TG-ReDial\tablefootnote{\url{https://github.com/RUCAIBox/TG-ReDial}} \\
Online Travel and Check-In & FourSquare\tablefootnote{\url{https://www.kaggle.com/datasets/chetanism/foursquare-nyc-and-tokyo-checkin-dataset}}, Fliggy\tablefootnote{\url{https://tianchi.aliyun.com/dataset/113649}}, Yelp\tablefootnote{\url{https://www.yelp.com/dataset}}, Gowalla\tablefootnote{\url{https://snap.stanford.edu/data/loc-Gowalla.html}}              \\
Online Shopping            & Amazon\tablefootnote{\url{https://cseweb.ucsd.edu/~jmcauley/datasets.html\#amazon_reviews}}, Taobao\tablefootnote{\url{https://tianchi.aliyun.com/dataset/649}}, RetailRocket\tablefootnote{\url{https://www.kaggle.com/datasets/retailrocket/ecommerce-dataset}}, Tafang\tablefootnote{\url{https://www.kaggle.com/datasets/chiranjivdas09/ta-feng-grocery-dataset}}  \\
Online Education           & MOOCCube\tablefootnote{\url{http://moocdata.cn/}}                            \\ \hline\hline
\end{tabular}
\label{tab:dataset}
\end{table}

\paratitle{Movie and Micro Video Recommendation.} In~\cite{zheng2024facetcrs, zheng2024hycorec, zheng2024mitigating, zheng2024diversity}, the authors extract the user multi-interest from item, entity-, word-, context-, knowledge-, and review-facet by a hypergraph neural network for conversational movie recommendation. CMI~\cite{li2022improving} adopts contrastive learning to improve the robustness of interest representations in micro-video recommendation.
In~\cite{jiang2020aspect}, the authors explicitly model the multi-scale time information and user interest group for micro video recommendation.

There are various public datasets available for movie and micro-video recommendations.
The \textit{MovieLens dataset}~\cite{harper2015movielens} contains user ratings and movie information (\eg movie genre) from the MovieLens website\footnote{\url{https://movielens.org/}}, available in multiple sizes (\eg 100K, 1M, 10M ratings).
The \textit{ReDial dataset}~\cite{li2018towards} consists of $10,006$ conversations about $51,699$ movies from IMDb. The \textit{TG-ReDial dataset}~\cite{zhou2020towards} is a Chinese dialogue dataset from Douban~\footnote{A Chinese online social networking service (\url{https://movie.douban.com/}) that enables registered users to record information and create content related to film, books, music, opera, and concerts.} with $10,000$ conversations and $129,392$ utterances about $33,834$ movies. All these datasets are widely used.

\paratitle{News Recommendation.} MINER~\cite{li2022miner} is a representative work in multi-interest recommendation, which encodes both news categories (\eg finance, sports, fitness, movies, etc) and contents for recommendation. Besides, a disagreement regularization term is incorporated into the loss function to enhance the disagreement of multi-interest representations. In~\cite{wang2021popularity},  the authors incorporate user intent regarding item popularity into multi-interest modeling, enhancing the recommendation performance for the hot news. 

\textit{MIND Dataset}~\cite{wu2019neural} is a popular news recommendation dataset, collected on the Microsoft News website. MIND includes approximately $160,000$ English news articles and over $15$ million impression logs generated by one million users. Each news article contains textual content, title, abstract, body, category, and entities. Each impression log records click events, non-clicked events, and the user's historical news click behavior before the current impression.

\paratitle{Life Services and Travel Recommendation.} Life service and travel recommendation aims to deliver travel-related products (\eg flight or train tickets, hotel bookings, package tours, etc) and local life-services (\eg dining, shopping, entertainment, etc) to customers, meeting their diverse needs. In~\cite{tao2022sminet}, the authors aim to model the user's spatial-temporal interest, group interest, and periodic interest from historical behaviors for check-in recommendation. M2GNN~\cite{huai2023m2gnn} considers the tags and cross-domain information with graph neural networks for life-service recommendation. 

Distinct from the datasets on other application tasks, time and spatial information are essential for check-in and travel-related recommendations. Some of the popular datasets in this category include: the \textit{FourSquare dataset} contains timestamped and geotagging check-in records from New York and Tokyo, revealing user activity patterns in location-based social networks; the \textit{Fliggy dataset}~\cite{tao2022sminet} is a large-scale travel platform dataset from Alibaba\footnote{Alibaba (\url{https://www.alibabagroup.com/en-US/}) is a Chinese multinational technology corporation with a specialization in e-commerce, retail, internet, and technology sectors. The company offers a comprehensive range of services, including consumer-to-consumer (C2C), business-to-consumer (B2C), and business-to-business (B2B) sales through both Chinese and global marketplaces. Additionally, Alibaba provides services in local consumer markets, digital media and entertainment, logistics, and cloud computing.}, including user behaviors and product information; the \textit{Yelp dataset} provides real-world business data such as reviews, photos, check-ins from the Yelp platform\footnote{Yelp (\url{https://www.yelp.com/}) is an American company that publishes crowd-sourced reviews about businesses.}; and the \textit{Gowalla dataset}~\cite{cho2011friendship} consists of over $6.4$ million check-ins from nearly $200,000$ users, recording location-based social network activity from 2009 to 2010.
    
\paratitle{Online Shopping Recommendation.} MIND is a pioneering work in multi-interest recommendation, which verifies the effectiveness of the proposed solution on a large-scale industrial dataset from Tmall.
On the light of MIND, ComiRec~\cite{cen2020controllable} and UMI~\cite{chai2022user} further apply multi-interest recommendation on the online platform. In~\cite{qin2022multi}, the authors devise a sophisticated model to capture users' diverse interests and the evolution of intention from both purchasing and browsing aspects. Besides, a real-world large-scale industrial dataset from the log data of Taobao~\footnote {Taobao is an online shopping platform from China, owned by Alibaba.} searching service is utilized for performance evaluation. In~\cite{chen2021exploring}, the authors consider the periodic and time interval information encapsulated in the sequential behaviors, thus, a graph neural network is applied for global and local information aggregation and multi-interest modeling. 
In~\cite{jiang2022adaptive}, the authors explore multi-interest modeling across multi-domains for display advertising product recommendation. 

The widely used datasets on online shopping include Amazon Review, Taobao, RetailRocket, and Ta Feng. The \textit{Amazon Review}~\cite{he2016ups, mcauley2015image, ni2019justifying, hou2024bridging} are among the most widely utilized datasets on various recommendation tasks. Collected from the Amazon platform, this dataset contains hundreds of millions of users, items, reviews, and rich side information (such as ratings, descriptions, prices, and images), with data spanning from May 1996 to September 2023. The \textit{Taobao dataset}~\cite{zhu2018learning, zhu2019joint, zhuo2020learning} focuses on user behaviors (\eg click, purchase, add to cart, and favor). It contains about one million users and four million items, collected over nine days in 2017, with each record detailing user/item IDs, category, behavior type, and timestamp. The \textit{RetailRocket dataset} is an e-commerce dataset comprising millions of user-item interactions gathered over $4.5$ months. The \textit{Ta Feng dataset} consists of four months of user shopping logs (from November 2000 to February 2001) from a Chinese grocery store, providing detailed records of users' purchase histories.

\paratitle{Online Course Recommendation.} Due to the elimination of time and location constraints, online education has become ubiquitous. Online education platforms, such as Massive Open Online Courses (MOOCs), provide a variety of courses for learners to select. In~\cite{liu2024personalized, yang2025explainable}, the authors apply attention and graph neural networks, respectively, for multi-interest modeling and online course recommendation. 

\textit{MOOCCouse}~\cite{yu2020mooccube} is a public course recommendation dataset collected from XuetangX\footnote{\url{https://next.xuetangx.com/}}, a prominent online education platform in China. These datasets encompass approximately one thousand courses, tens of thousands of learners, and learner-user interaction records for recommendation.

\section{Challenges and Future Directions}
\label{sec:challenges}

Despite the widespread adoption and a long track of research in multi-interest recommendation, there remain several challenges and pivotal directions where existing works have yet to be fully addressed and explored. In this section, we outline the following prospective and promising research directions, which are critically important for further development in multi-interest recommendation. 

\subsection{Adaptive Multi-Interest Extraction}

Reviewing existing works, it is noteworthy that most solutions require predefining a fixed interest number for multi-interest extraction, as introduced in Section~\ref{sec:ext}. However, given the diversity and complexity of user behaviors and domains, we argue it is inappropriate and impractical in real-world scenarios. For instance, in the context of news recommendations, it might be feasible to predefine a fixed number of multi-interests based on topics (\eg sports, finance, and policy), as the topics of most news articles are generally straightforward and singular in practice. In contrast, this approach becomes intractable for literature recommendations due to the inherent uncertainty, ambiguity of themes, language styles, and underlying thoughts. Additionally, users may also exhibit varying behavioral preferences in the real world, \ie some individuals may maintain consistent shopping habits over time, while others may shift their interests in response to changes in personal environments.
Moreover, the fixed number of multiple interests impedes the knowledge transfer across different domains and further compromises the performance of the final recommendations.     
Therefore, adaptive multi-interest extraction is rational for addressing the complexity of practical scenarios. 
In~\cite{tang2023towards, huhorae}, the authors propose a sparse interest capsule activation function, which adaptively activates the most relevant interest captures by calculating the similarity between the representation of items and interests, thereby allowing an adaptive multi-interest modeling.
Inspired by some sophisticated techniques, we can also apply density-based clustering methods (\eg DBSCANE~\cite{martin1996dbscane}), which analyze data density to determine the multi-interest numbers automatically. Besides, hierarchical clustering methods~\cite{ran2023cluster} can also be utilized, as these models construct dendrograms to split or combine clusters based on branch shapes, allowing for dynamic adjustment of cluster numbers. Furthermore, we can calculate the silhouette score~\cite{ketan2020silhouette} (\ie measure the intra-cluster cohesion and inter-cluster separation of each sample) or gap statistic~\cite{tibshirani2001gap} (\ie compare observed within-cluster variance to null distributions) to obtain the optimal number of clusters for multi-interest extraction.

\subsection{Efficiency in Multi-Interest Modeling}

Owing to the multi-interest extraction and modeling, the computational costs associated with multi-interest recommendation are significantly higher compared to traditional recommendation solutions.
Therefore, the efficiency of multi-interest recommendation has been met with concerns regarding resource efficiency and practical expense. To be specific, the primary bottlenecks affecting training efficiency stem from the following aspects.

\paratitle{Efficiency in Multiple Number of Interest Representations Modeling.} 
As discussed in~\cite{li2019multi, wang2022target}, there exists a trade-off between recommendation performance and training efficiency with regard to the number of interests. Specifically, when the number of interests is insufficient, the representational capacity of models will be compromised, hurting the final recommendation results. 
Conversely, an excessively large number of interests significantly increases both the computational and memory footprint. Therefore, a proper number of interests is crucial for an effective and efficient recommendation.

\paratitle{Efficiency in Iteratively Multi-Interest Extraction.} The majority of existing works employ dynamic routing for multi-interest extraction, which relies on an iterative agreement process to update coupling coefficients between capsules. Compared to the other methods, such as attention (\ie only requires a single forward pass for multi-interest extraction), this approach caused increased computational time proportional to the number of iterations. 
Additionally, the number of learnable matrices between capsules increases quadratically with the number of capsules (\ie $t\times K$ learnable weight matrices in Eq.~\eqref{eq:dynamic_routing}). This property incurs high computational and memory costs for large-scale and complex interactions, hindering further deployment in practical scenarios.

\paratitle{Efficiency in External Side Information Modeling.} As introduced in Section~\ref{sec:diemnsion}, explicit multi-interest modeling depends on the side information, such as user behaviors, spatial-temporal information, item attributes, and multi-modality information. However, encoding this external information requires dedicated modules, which introduce extra learnable parameters. This issue becomes more pronounced when large language models or large vision models are engaged.
In addition, not all the side information is universally beneficial across different recommendation scenarios. For example, for online travel and check-in recommendations, the spatial and temporal information is crucial, whereas the genres and user profile are more critical for movie and book recommendations.
Consequently, identifying the types of side information is truly essential for both recommendation performance and efficiency improvement.
    
\paratitle{Efficiency in Multi-Interest Representation Aggregation.} There are two paradigms for multi-interest aggregation, \ie multi-interest representation aggregation and multi-interest recommendation aggregation, as outlined in Section~\ref{sec:agg}. Each of these strategies has its advantages and disadvantages. Although aggregation on recommendation results is more straightforward and avoids the introduction of additional learnable parameters, this operation incurs extensive computational cost when the candidate items are extremely large. On the other hand, representation fusion demonstrates more efficiency when calculating prediction scores, but the tailored representation aggregation modules introduce additional overhead. Therefore, the consideration of the aggregation method warrants detailed discussion in practical scenarios.

\subsection{Multi-Interest Extraction for Denoising} 
Existing approaches often consider all items for interest modeling, even though not all items are pertinent to the user's current intention; some may even be noise due to user misoperations. Additionally, item-associated information, especially for descriptions, contains substantial noise. This misinformation will mislead the recommender system, diminishing the final performance~\cite{wang2021denoising, qin2021world}. 
In multi-interest recommendation, techniques such as dynamic routing or attention mechanism in the multi-interest extractor can be recognized as a soft selection strategy. As a result, the most relevant information and features are selected, while the noise information will be diminished for the recommendation.
However, relatively few works on multi-interest modeling have explored the contribution of denoising to the recommendation results.
We believe that this research direction deserves further in-depth exploration.

\subsection{Explainability in Multi-Interest to Multi-Aspect Alignment}
Beyond the accuracy and efficiency of results, the explainability of recommendations is also vital for an advanced recommendation system. 
An explainable recommendation can further improve the transparency and persuasiveness of systems, boosting the satisfaction and stickiness of users~\cite{zhang2020explainable}. 
Existing research regarding the explainable recommendation can be categorized into four methodologies~\cite{chatti2024visualization}: (1) content-based~\cite{de2015semantics, li2025efficient}, (2) collaborative-based~\cite{balog2019transparent, koren2021advances}, (3) social-based~\cite{sharma2013social}, and (4) hybrid methods~\cite{kouki2019personalized}.
More than conventional explainable recommendation requirements, the explainability of multi-interest recommendation aims to provide a more in-depth analysis of users' tendencies towards each interest and the connection of each item to these interests by exploring the underlying intentions behind the interactions.
For instance, in~\cite{li2019capsule}, the authors jointly consider the item aspects and the associated viewpoints from users by analyzing the sentiment of reviews for explainable recommendation. To this end, a capsule neural network was applied to estimate the effects of each aspect on sentiment.
Multi-interest modeling imposes greater demands and additional challenges for explainable recommendation systems. 
However, this area has received limited attention in existing research. It holds significant promise for future exploration.

\subsection{Multi-Interest Alleviate Long Tail and Cold-Start Problem}
Both the long tail\footnote{A small group of items garners the majority of interaction records and dominates the recommendations, leaving a large number of items with insufficient exposure opportunities.} and cold-start~\footnote{Due to a lack of sufficient historical interaction records, it is challenging for new users or items to make accurate recommendations.} problems are persistent challenges~\cite{yin2021tail, Li2017tail, zhang2025cold} and significantly undermines the fairness and diversity of recommendation systems. 
As the cause of these two problems lies in data sparsity, the basic idea of the fundamental solution is to introduce more external information for data and representation augmentation.
Multi-interest recommendation attempts to capture the user's multiple interests towards the items' multiple aspects by leveraging such external information, thus, it exhibits great potential for alleviating the tail item recommendation problem. Specifically, auxiliary knowledge has been leveraged in multi-interest modeling to address challenges of cold-start and tail item recommendation. This includes cross-domain information~\cite{sun2023remit, huai2023m2gnn, zhang2022diverse, zang2023contrastive}, multi-modal information~\cite{wang2023missrec, li2025reembedding}, spatial-temporal and user group information~\cite{tao2022sminet}, as well as co-occurring head items~\cite{liu2023co}.


\subsection{Frontier Methodology in Multi-Interest Recommendation}

While multi-interest recommendation systems have been extensively studied by both academia and industry, several frontier research directions remain to be explored further.

\paratitle{Reinforcement Learning in Multi-Interest Recommendation.}
Different from supervised learning approaches, reinforcement learning~\cite{kaelbling1996reinforcement, wang2022deep} can be regarded as a goal-oriented algorithm where an agent learns to maximize cumulative rewards through environmental interactions. 
By framing user interaction behaviors as real-time decision-making processes, reinforcement learning facilitates users' dynamic interests and interest evolution modeling in recommendation.
Recent advancements have demonstrated reinforcement learning's growing effectiveness in recommendation systems~\cite{afsar2022reinforcement, lin2023survey}. 
As to the multi-interest recommendation, reinforcement learning can be applied for multi-interest or candidate item selection. 
Specifically, in~\cite{sun2023remit}, the authors consider user multi-interest for the state modeling, wherein the action is defined as selecting which specific interest will be activated for the currently recommended items. Recommendation accuracy is treated as the reward value for policy updates.
In contrast, in~\cite{zhang2022multiple, shen2024multi}, the authors define the actions as candidate item selection based on multi-interest modeling.
We believe reinforcement learning can be used to balance the trade-off between exploration and exploitation of each interest, ensuring that the system not only capitalizes on known user interests but also discovers the user's potential multiple interests, enhancing the diversity of recommendations.

\paratitle{Large Language and Vision Models in Multi-Interest Recommendation.}
Encouraged by the remarkable success of large language and vision models in a wide spectrum of downstream tasks~\cite{thirunavukarasu2023large, ren2024survey, zan2023large, chen2025large, tang2024text}, large language models and multi-modality recommendation have emerged as powerful methods for multi-interest recommendation~\cite{tangone, zhao2024recommender, wen2024unified, li2025efficient}.
In~\cite{li2022miner}, the authors apply a language model for news content encoding, which will be further fed into a multi-interest extractor for multi-interest representation generation.  
In~\cite{tang2023towards}, the authors propose a multi-interest pre-training framework, which first applies a large language model for item contextual information modeling. Besides, a sparse capsule network is proposed to adaptively extract multiple user interests.
Building upon this foundation, \cite{huhorae} further incorporates the relative position information and the relative time interval information jointly when performing multi-interest learning.
Considering the inherent gap between the specific domain knowledge of recommendations and world knowledge in large language models, in~\cite{bai2024aligning}, the authors extend the direct preference optimization (DPO) algorithm~\cite{rafailov2023direct} to align the large language models with user multi-preference modeling for recommendation.
In~\cite{wang2023missrec}, the authors propose a multi-modal pre-training framework to extract the user's multi-interest from different modality information for recommendation. 
In contrast to large language models, few studies concentrated on multi-modal and multi-interest recommendations. 
Attributing to the fertile side information available in multi-modal data, we believe this research area will propel the multi-interest recommendation into a new frontier.
 
\paratitle{Diffusion Models in Multi-Interest Recommendation.} 
In contrast to the auto-regressive generation paradigm used by large language models, diffusion models introduce a novel framework for generative tasks, demonstrating exceptional performance across diverse applications~\cite{lin2024survey, li2023diffurec, yang2024survey}. Essentially, diffusion models operate in two phases: a forward diffusion process, which progressively degrades the original input into Gaussian noise, then a reverse process, which reconstructs the data step-by-step from the noise, conditioned on the auxiliary information from the input. 
For multi-interest recommendation, diffusion methods are not fully exploited. In~\cite {li2024multi, le2025diffusion}, the authors leverage the reverse process of diffusion models to mitigate the impact of noisy interactions and obtain refined multi-interest representations for recommendations.
However, we argue that applying the diffusion model to recover users’ true multi-faceted interests accurately is challenging, given the inherent ambiguity and uncertainty of user interests, as well as the incompleteness and bias of historical interaction data caused by the exposure mechanisms of recommender systems~\cite{benigni2025diffusion}.
Overall, further efforts are required to effectively apply diffusion methods for diversity and uncertainty modeling, as well as for multi-interest recommendation.

\section{Conclusion}
\label{sec:conclu}

This survey comprehensively investigated multi-interest modeling in recommendation systems. Specifically, we first reviewed over $170$ related works, introducing the origins, developments, and remarkable progress in this research area. Besides, we formalized the multi-interest modeling in recommendation tasks, clarified the main concepts, and categorized the aspects of multi-interest modeling from both the item-side and the user-side. 
Furthermore, a systematic classification was provided for existing work organizations based on their focusing tasks, multi-interest modeling aspects, and research concerns. 
We illustrated an overall framework of multi-interest recommendation, thus, the technical details, including the multi-interest extractor and multi-interest aggregation, were further introduced. 
The application scenarios and publicly available datasets were also provided. 
Finally, we summarized the main challenges and future promising directions in this area. We hope this survey offers readers a foundational understanding of multi-interest recommendation and inspires future research endeavors.

\begin{acks}
This work was supported by National Natural Science Foundation of China (No. 62272349, No. U23A20305, and No. 62302345); Natural Science Foundation of Hubei Province under Grant Numbers 2023BAB160. Chenliang Li is the corresponding author.
\end{acks}

\bibliographystyle{ACM-Reference-Format}
\bibliography{ref}

\appendix

\end{document}